\documentclass[iop]{emulateapj}
\usepackage{amssymb}
\usepackage{natbib}
\usepackage{amsmath}
\usepackage{subfigure}
\usepackage{footnote}
\usepackage{color}
\usepackage{mdwlist}
\usepackage{enumerate}
\usepackage{graphicx}
\newcommand{\siiv}{{\ion{Si}{4}}}

\newcommand{\fexxi}{{\ion{Fe}{21}}}
\newcommand{\fexxiv}{{\ion{Fe}{24}}}
\newcommand{\fexviii}{{\ion{Fe}{18}}}

\newcommand{\mgii}{{\ion{Mg}{2}}}
\def\arcsec{$^{\prime\prime}$}

\begin{document}

\title{Investigating the response of loop plasma to nanoflare heating using RADYN simulations}
\author{V.Polito}
\author{P. Testa}
\affil{Harvard-Smithsonian Center for Astrophysics, 60 Garden Street, Cambridge MA 01238, USA}
\author{J. Allred}
\affil{NASA Goddard Space Flight Center, Greenbelt, MD 20771, United States}
\author{B. De Pontieu}
\affil{Lockheed Martin Solar \& Astrophysics Lab, Org. A021S, Bldg. 252, 3251 Hanover Street Palo Alto, CA 94304, USA}
\affil{Rosseland Centre for Solar Physics, University of Oslo, P.O. Box 1029
Blindern, NO-0315 Oslo, Norway}
\affil{Institute of Theoretical Astrophysics, University of Oslo, P.O. Box 1029, Blindern NO-0315, Oslo, Norway}
\author{M. Carlsson}
\author{T. M. D. Pereira}
\affil{Rosseland Centre for Solar Physics, University of Oslo, P.O. Box 1029
Blindern, NO-0315 Oslo, Norway}
\affil{Institute of Theoretical Astrophysics, University of Oslo, P.O. Box 1029, Blindern NO-0315, Oslo, Norway}
\author{Milan Go\v{s}i\'c}
\affil{Lockheed Martin Solar \& Astrophysics Lab, Org. A021S, Bldg. 252, 3251 Hanover Street Palo Alto, CA 94304, USA}
\affil{Bay Area Environmental Research Institute, Petaluma, CA 94952, USA}
\author{Fabio Reale}
\affil{Dipartimento di Fisica e Chimica, Universit\`a di Palermo and Istituto Nazionale di Astrofisica (INAF)/Osservatorio Astronomico di Palermo, Piazza del Parlamento 1, 90134 Palermo, Italy}

\begin{abstract}
We present the results of 1D hydrodynamic simulations of coronal loops which are subject to nanoflares, caused by either in-situ thermal heating, or non-thermal electrons (NTE) beams. The synthesized intensity and Doppler shifts can be directly compared with IRIS and AIA observations of rapid variability in the transition region (TR) of coronal loops, associated with transient coronal heating. We find that NTE with high enough low-energy cutoff (E$_\textrm{C}$) deposit energy in the lower TR and chromosphere causing blueshifts (up to~$\sim$~20 km/s) in the \emph{IRIS} \siiv~lines, which thermal conduction cannot reproduce. The E$_\textrm{C}$ threshold value for the blueshifts depends on the total energy of the events ($\approx$~5 keV for 10$^{24}$ ergs, up to 15 keV for 10$^{25}$ ergs).   The observed footpoint emission intensity and flows, combined with the simulations, can provide constraints on both the energy of the heating event and E$_\textrm{C}$.  The response of the loop plasma to nanoflares depends crucially on the electron density: significant \siiv~intensity enhancements and flows are observed only for initially low-density loops  ($<$~10$^{9}$~cm$^{-3}$). This provides a possible explanation of the relative scarcity of observations of significant moss variability. While the TR response to single heating episodes can be clearly observed, the predicted coronal emission (AIA 94\AA) for single strands is below current detectability, and can only be observed when several strands are heated closely in time. Finally, we show that the analysis of the IRIS \mgii~chromospheric lines can help further constrain the properties of the heating mechanisms.

\end{abstract}

\section{Introduction}
Understanding the details of the physical mechanism(s) responsible for the heating of the solar and stellar coronae to temperatures exceeding several million Kelvin still represents one of the most challenging problems in astrophysics. Although significant progress in theoretical modeling and major observational advancements
have been achieved in recent years, we still lack a definitive solution to the coronal heating problem \citep[e.g. see reviews by][]{Klimchuk06,Klimchuk15,Reale14, Testa15}.
Among different competing mechanisms, small-scale impulsive nanoflare events (E $\approx$ 10$^{24}$--10$^{25}$ ergs), associated with magnetic reconnection in the corona, are  a popular candidate process for converting magnetic energy into the thermal energy required to heat the active region (AR) plasma \citep[e.g.][]{Parker88,Priest02}. We note that coronal heating models alternative to the nanoflare scenario, such as heating by Alfv\'en waves and associated with spicules, have also been suggested \citep[e.g.][]{vanBallegooijen11,DePontieu11,DePontieu17}. 

Several authors have focused on constraining the properties of the coronal heating models by comparing
the plasma parameters  derived from spectroscopic and imaging observations (i.e. flows, line broadening, emission measure, line intensity ratio) with the predictions of numerical simulations \citep[e.g.][]{Warren08,Reale09,Winebarger11,Testa11,Testa16}. 
However, observing the direct signatures of energy release in the corona is complicated by many factors, including the low-density of the plasma and the presence of non-equilibrium effects, and therefore only limited constraints can often be obtained for the models. 

On the other hand, the observation of the upper transition region (TR; T $\sim$ 0.08-0.15 MK) footpoints of hot (T $\gtrsim$ 3MK) coronal loops (i.e. "moss") provides highly sensitive diagnostics, because of the rapid variation of the physical conditions of the local plasma in response to the heating. The low-lying coronal plasma was first obtained in the X-rays with the \emph{Normal Incidence X-Ray Telescope} \citep[NIXT; e.g.][]{Peres94}, and then identified as upper TR moss by the \emph{Transition Region and Coronal Explorer} \citep[\emph{TRACE}; ][]{Handy99} in narrow EUV bands centered around $\approx$~1~MK \citep[e.g.][]{Fletcher99,Berger99,DePontieu99,Martens2000}. Some of the studies using \emph{TRACE} images, as well as spectroscopic observations from instruments such as the \emph{Hinode}/EUV Imaging Spectrometer \citep[EIS; ][]{Culhane07}, showed low variability of the moss region, on timescales of the order of several minutes, which had often been interpreted as an indication of quasi-steady heating of the coronal loops \citep[e.g.][]{Antiochos03,Brooks09,Warren10,Tripathi10}, or as due to other phenomena \citep[e.g.][who established that variability on timescales of 3-5 min is caused by chromospheric jets obscuring upper TR material]{DePontieu03}. In 2012, the launch of the \emph{High-resolution Coronal Imager} \citep[\emph{Hi-C}; ][]{Kobayashi14}, on board a sounding rocket flight, provided a very different and much clearer view of the moss dynamics, thanks to its unprecedented spatial (0.3\arcsec--0.4\arcsec) and temporal (5.5~s) resolution.~\emph{Hi-C} observations revealed highly variable TR moss emission on timescales down to 15~s associated with coronal activity, supporting the impulsive nanoflare heating scenario \citep[e.g.][]{Testa13}.  However, \emph{Hi-C} lacked spectroscopic measurements, which are crucial to provide detailed information about the physical conditions of the plasma and therefore key observables to compare with the predictions of the models. 

Since its launch in 2013, the
\emph{Interface Region Imaging Spectrometer} \citep[\emph{IRIS}; ][]{DePontieu14} has provided simultaneous observations of chromosphere and lower TR spectra and images at very high spatial (0.33\arcsec), spectral ($\approx$~3 km$\cdot$s$^{-1}$) and temporal (up to 1--2s) resolutions. \emph{IRIS} has greatly improved on the past spectroscopic instruments, which were limited by lower spatial resolution and the lack of co-aligned context imaging observations. 
In particular, recent work by \citet[hereafter T14]{Testa14}  has demonstrated that \emph{IRIS} observations of the TR footpoint variability combined with advanced numerical simulations can provide tight constraints on the nanoflare models. More specifically, T14 showed that the blueshifts observed in the TR \siiv~line at 1402.77~\AA~(T$\approx$~10$^{4.9}$ K) with \emph{IRIS} could not be reproduced by models assuming heating by thermal conduction only, but were consistent with electron beam heating, highlighting for the first time the possible importance of non-thermal electrons in the heating of non-flaring ARs.  T14 also combined images from the Atmospheric Imaging Assembly \citep[AIA; ][]{Lemen12} on board the \emph{Solar Dynamic Observatory} \citep[\emph{SDO}; ][]{Pesnell12} to precisely locate the TR footpoints of the hot coronal loops. \citet{Judge17} recently analyzed additional magnetic data for the observations of T14 and found that at one location the ribbon appears possibly magnetically disconnected from the corona. In light of this intriguing possibility, \citet{Judge17}  discuss alternative heating scenarios (e.g., chromospheric reconnection) for that one ribbon location, although, as already discussed by T14, these alternative mechanisms appear unlikely for the vast majority of the observed TR brightenings. 

The numerical investigation carried out by T14 was performed with the RADYN code \citep{Carlsson92, Carlsson95, Carlsson97, Allred15} using a limited set of model input parameters aimed at reproducing the particular observation under study. However, \emph{IRIS} observations indicate a large variety of properties for the observed footpoints brightenings, which show lifetimes of 10--30 s, a broad range of Doppler shifts in the TR and chromospheric lines (from blue to red or no-shifts) and appear connected to coronal loops with a range of different lengths ($\sim$~10--100 Mm), as observed by AIA (Testa et al., in prep). To understand and interpret the variety of TR observations, in this work we carry out a detailed analysis of nanoflare simulations, exploring a much wider region in the parameter space compared with the analysis of T14. In particular, we explore the atmospheric response to different nanoflare heating models as a function of parameters such as  the electron low energy cut-off, initial loop temperature, length and total nanoflare energy. We also compare in detail the predictions of nanoflare models assuming heating by either electron beams or in-situ heating of the corona.   Here we do not aim to forward model a specific observation, but we wish to provide a grid of models, with a wide range of parameter values, that can be used as a guide for the interpretation of  \emph{IRIS} observations of footpoint variability associated with coronal heating. At the same time, we aim to investigate the applicability and limitations of the \emph{IRIS} and \emph{SDO}/AIA diagnostics of nanoflare heating. 

The paper is organized as follows: in Sect.~\ref{Sect:2}  we describe the RADYN code and the set of nanoflare simulations we performed. Section~\ref{Sect:3} presents the evolution of atmospheric response for one set of simulations, whereas Sects.~\ref{Sect:4} examines the forward modelling of the \emph{IRIS} and AIA optically thin plasma observables. In Sect.~\ref{Sect:5}  we briefly present the results of synthethic chromospheric \mgii~emission observed by \emph{IRIS} for our different nanoflare models.The main results for the whole parameter space are then discussed in Sect.~\ref{Sect:6}. Finally, in Sect.~\ref{Sect:7} we summarize our main findings and draw our conclusions.

\section{The numerical investigation}
\label{Sect:2}
We perform simulations of nanoflare-heated loops using the RADYN numerical code. RADYN solves the equation of charge conservation and the level population rate equations on a  single 1D magnetic strand in the field-aligned direction, rooted in the photosphere and stretching out to include the chromosphere, TR and corona.
A key advantage of RADYN compared to other 1D hydrodynamic codes is the ability to model the important elements for the chromospheric energy balance (H, He, Ca II) in non-local thermodynamic equilibrium (non-LTE). The chromospheric losses for these elements are calculated in detail using non-LTE radiative transfer, whereas other atomic species are included as background continuum opacity source (in LTE) using the Uppsala opacity package \citep{Gustafsson73}. The radiative loss functions for optically thin lines are calculated using the CHIANTI 7.1 \citep{Dere97,Landi13} atomic database assuming ionization and thermal equilibrium. RADYN uses an adaptive grid \citep{Dorfi87} with a fixed number of grid cells, the size and location of which can vary to allow shocks and steep gradients in the atmosphere to be resolved. 

In this work, we use the state-of-the-art version of RADYN described in \cite{Allred15}, with the following two main modifications: (1) we extended the number of grid points (from 191 to 300), to allow for a sufficient resolution of even the steepest gradients arising from the release of energy in the loop;  (2) we added a more realistic atmospheric structure for the loops, including a plage-like chromosphere following the work of \cite{Carlsson15}. 

As discussed in the introduction, there is observational evidence of high variability in the TR and chromosphere associated with coronal heating, which, combined with modeling, can provide important diagnostics of the heating properties. Here we aim to use RADYN to explore two likely scenarios for coronal nanoflares models: (1) transport by non-thermal electrons (accelerated in the corona) which deposit their energy in different layers of the atmosphere, depending on their energy distribution properties and the physical conditions of the plasma in the loop. A power-law distribution for the electron beams is usually assumed in RADYN, with user-specified values of low energy cut off (E$_\textrm{C}$; keV), total energy (E; erg), energy flux (F; erg cm$^{-2}$ s$^{-1}$) and spectral index ($\delta$); (2) local heating of the corona, with subsequent transport of the energy to the lower atmosphere by thermal conduction. 
To this aim, we have explored an extensive parameter space for our simulations guided by previous modeling results (T14) and recent observations (Testa et al. in prep), including different heating and loop properties. In particular, as summarized and discussed below:

\begin{enumerate} 
\item Heating models : 
\begin{enumerate} 
\item Electron beam (hereafter, EB) model, with E~=~6$\cdot$10$^{24}$ erg, $\delta$ = 7 and E$_\textrm{C}$~=~5, 10 and 15 keV

\item In-situ heating of the corona at the loop top (thermal conduction model, hereafter TC), with total energy of 6$\cdot$10$^{24}$ erg
\end{enumerate}
\item Half-loop lengths \emph{L}/2: 15 Mm, 50 Mm and 100 Mm
\item Initial loop-top temperatures (T$_{\textrm{LT}}$): 1 MK, 3 MK and 5 MK
\end{enumerate}
The total energy that we choose to deposit in a single loop is representative of the typical values assumed for nanoflares \citep[10$^{24}$--10$^{25}$ erg, e.g.][]{Parker88}, and it reproduced TR brightenings similar to the \emph{IRIS} observations in T14, whereas the choice of  $\delta$ = 7 for the electron energy distribution is motivated by the observational trend of increasing spectral index with decreasing flare class \citep[e.g.][]{Hannah11}. A steep electron distribution means that more electrons have energy close to E$_\textrm{C}$ and are therefore closer to iso-energetic beams, allowing us to study separately the effects of energy deposition by electrons of varying energies. We choose 3 values of low energy cut-off E$_\textrm{C}$, resulting in beams which are dominated by low (5 keV), intermediate (10 keV) and high (15 keV) energy electrons.

Our initial loops are obtained assuming initial hydrostatic equilibrium for the plasma, and therefore loops at different temperatures will have different densities. In particular, hotter loops will also be denser. We chose a range of different initial temperature and density conditions which aim to reproduce the large variety of observed active region loops at different heating stages. In particular, low-temperature and longer loops have a lower density and are representative of empty strands before any heating has taken place, whereas hotter (and denser) loops represent strands which have been previously heated and filled with plasma. 

All the simulations have been performed for a loop with a cross-sectional area \emph{A} of 5$\cdot$10$^{14}$ cm$^{2}$ (corresponding to a diameter of $\approx$~250~km). 
Note that the loop cross-section area does not enter directly in the model, but is only a normalization parameter. We assume the cross-section to be constant in our calculations. However, \cite{Mikic13} showed that the assumption of a non-uniform cross-section might affect the evolution of the loops and, in particular,  enhance the coronal emission, as we also briefly discuss in Sect.~\ref{Sect:4.2}.


Further, we assume that the loop strand is heated constantly for 10~s, which is consistent with the lifetimes of short-lived brightenings typically observed in TR moss, as discussed in the introduction. Given the values of loop cross-section and heating duration above, the electron beam energy flux (F) will be of the order of 1.2$\cdot$10$^{9}$ erg s$^{-1}$ cm$^{-2}$ for our EB model runs. 
Finally, we also discuss the effect of varying the total energy input (from 10$^{24}$ to 10$^{25}$ erg) for one set of loop length and initial apex temperature. 

\begin{figure}
\centering
\includegraphics[width=0.45\textwidth]{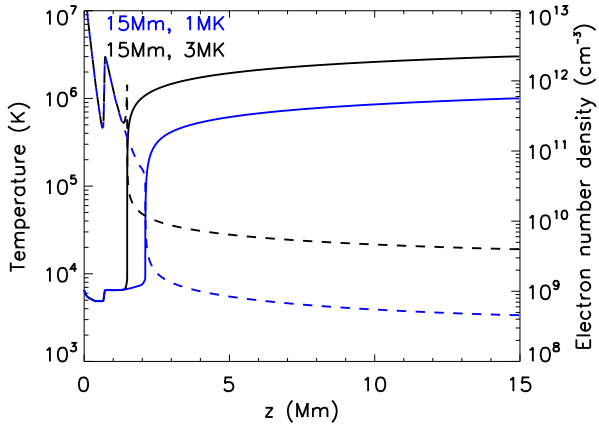}

\caption{Temperature (continuous lines) and electron number density (dotted lines) for the loops with plage-like atmospheres and initial apex temperature of 1~MK (blue) and 3~MK (black).}
\label{Fig:init_atm}

\end{figure}

As mentioned above, we adopt an initial atmosphere for the 1D loops that is based on the work by \cite{Carlsson15}, aimed at reproducing  \mgii~line profiles which are closer to the observed ones in the plage.  Figure~\ref{Fig:init_atm} shows the initial atmospheric structure for the 15~Mm loop with initial apex temperatures of 1~MK and 3~MK that we adopted in this work. Note the presence of a hot chromosphere with a steep temperature rise. 

Over 40 numerical simulations were performed and our key findings are presented as follows: in Sect.~\ref{Sect:3} we discuss the results of atmospheric response of the loop to the nanoflare heating for the model run with 15 Mm loop and initial T$_{\textrm{LT}}$ of 1 MK and 3 MK, and in Sects.~\ref{Sect:4} and \ref{Sect:5} we describe the model predictions of specific \emph{IRIS} and AIA observables and associated diagnostics for these sets of simulations. In Sect. \ref{Sect:6}, the results for the full parameter space, including other loop lengths (50 and 100 Mm), temperatures (5MK), and different total energy input (E=10$^{24}$--10$^{25}$ erg) will be briefly discussed, mainly focusing on the differences with the model predictions presented in Sects.~\ref{Sect:3}, \ref{Sect:4} and \ref{Sect:5}.
Finally, Sect.~\ref{Sect:7} summarizes the key conclusions of our numerical investigation.
\begin{figure*}
\centering
\includegraphics[width=\textwidth]{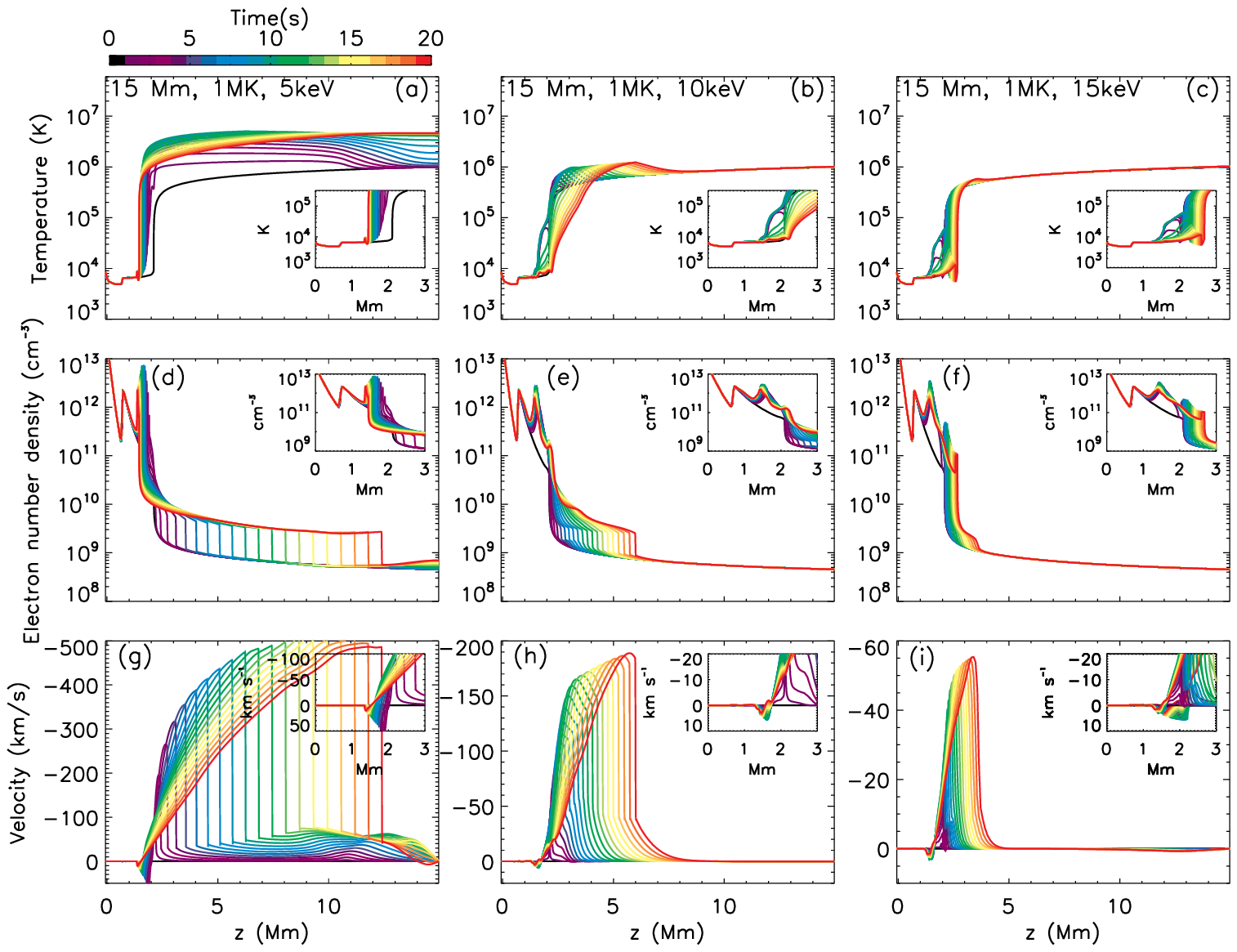}
\includegraphics[width=\textwidth]{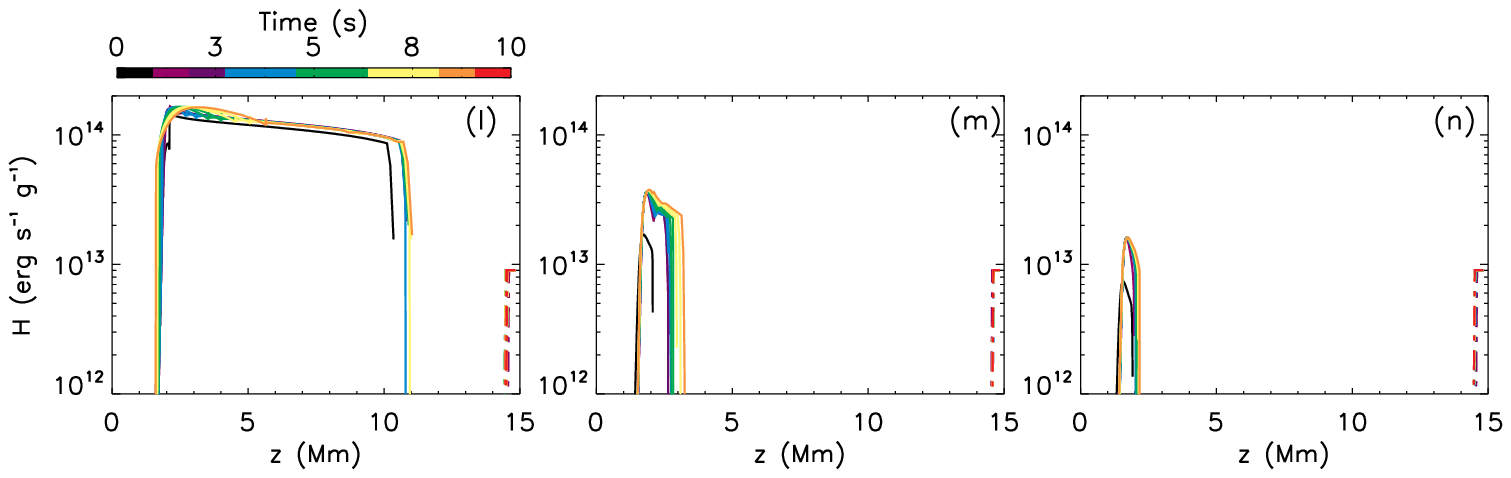}
\caption{Atmospheric response to the RADYN nanoflare simulations for the 15 Mm loop with initial loop top temperature of 1MK. From top to bottom: temperature, electron density, velocity and heating deposition rate per mass as a function of time for EB simulations with E$_\textrm{C}$ of 5 keV (left), 10 keV (middle) and 15 keV (right). Color indicates different times as shown in the colorbars, with the simulation output plotted every 1s. Negative/positive velocities indicate blueshift/redshift of plasma.  For panels a--i, the inserts show a zoom of the atmospheric evolution of the region \emph{z}=0--3~Mm. The different panels show (from top to bottom): the plasma temperature, electron density,  bulk velocity and heating rate per mass (\emph{H}). The dotted lines in panels \emph{l} to \emph{n} show the background heating per mass applied to maintain the initial steady loop atmosphere at the chosen pre-nanoflare temperature. Note that the heating rate is plotted with a different colorscale (with relative colorbar) as the heating is switched off after 10~s in the simulations.} 
\label{Fig:atm_resp_15Mm_1MK}

\end{figure*}

\section{Atmospheric response for the 15 Mm loop}
\label{Sect:3}
In this section, we describe the evolution of the atmospheric variables as a function of time for the 15 Mm half-loop length models, assuming electron beam (Sect.~\ref{Sect:3.1}) and in-situ (Sect. \ref{Sect:3.2}) heating. 

\subsection{Electron beam heating}
\label{Sect:3.1}
\begin{figure*}
\centering
\includegraphics[width=\textwidth]{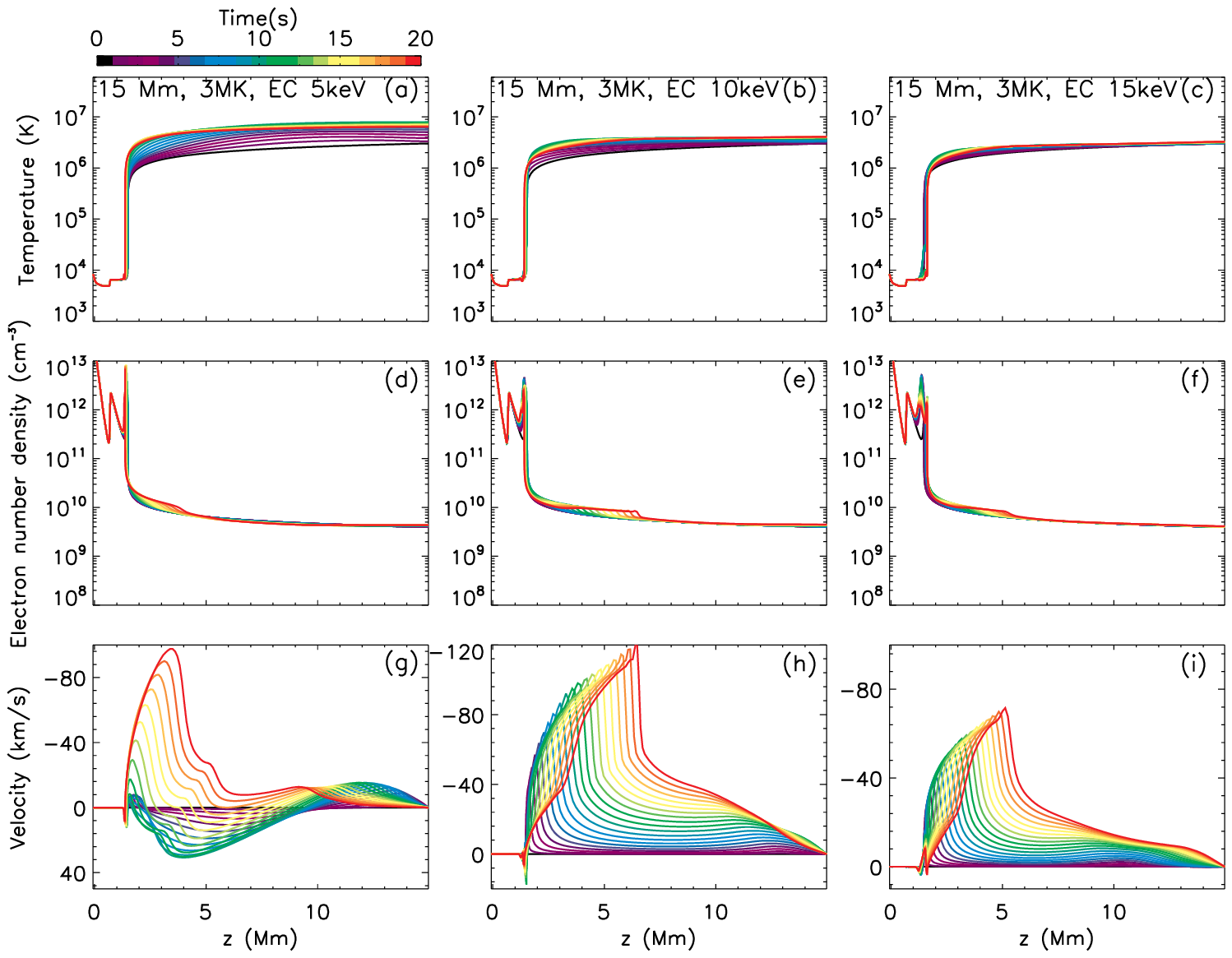}
\includegraphics[width=\textwidth]{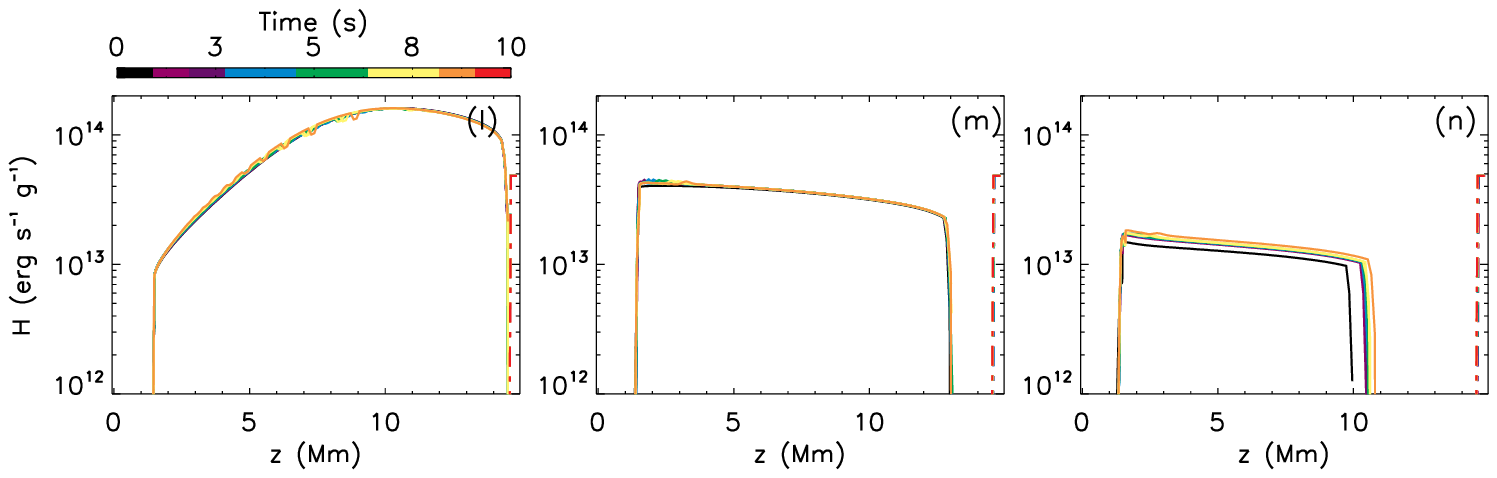}
\caption{Atmospheric response to the RADYN nanoflare simulations for the 15 Mm loop with initial loop top temperature of 3~MK. See the caption of Fig.~\ref{Fig:atm_resp_15Mm_3MK} for an explanation of the different panels. }
\label{Fig:atm_resp_15Mm_3MK}
\end{figure*}

Figures~\ref{Fig:atm_resp_15Mm_1MK}  and \ref{Fig:atm_resp_15Mm_3MK}  compare the evolution of the atmospheric variables in the EB heating models as a function of different E$_\textrm{C}$ for the 15 Mm half-loop with initial loop-top temperature T$_{\textrm{LT}}$~=~1 and 3~MK. These two loops have coronal density N$_{e,LT}$~=~$\approx$~10$^{8.7}$ and 10$^{9.6}$~cm$^{-3}$ respectively.  Both figures show (from top to bottom): the plasma temperature, electron density, velocity and beam heating deposition rate per mass \emph{H} as a function of time for the first 20 s into the simulations with a 1s interval. \emph{H} is given by the ratio of the particle beam heating rate \emph{Q$_\textrm{beam}$} \cite[see Eq. 18 of ][]{Allred15} and the plasma density. Note that only the first 10~s of the simulation are shown in the heating rate plots as the heating is switched off after this time. We focus here on the discussion of the first 20~s, because we are interested in comparing in detail the atmospheric response for different models during the heating and the early relaxation phase after the heating is switched off. We point out that the evolution timescales will be different for loops with different temperatures and densities. Nevertheless, we focus our attention on a small timescale because, as mentioned earlier, the lifetimes of the observed TR moss brightenings are of the orders of 10--30~s \citep[][Testa et al. in prep]{Testa13, Testa14}.

The different columns  in Figs.~\ref{Fig:atm_resp_15Mm_1MK}  and \ref{Fig:atm_resp_15Mm_3MK}  show the results for EB simulations with (from left to right): E$_\textrm{C}$ = 5, 10 and 15 keV respectively. Note that negative (positive) velocities mean upflows (downflows) of plasma. The inserts in Fig.~\ref{Fig:atm_resp_15Mm_1MK} show a zoom in the dynamics of the chromospheric evolution between \emph{z}~=~1--3~Mm. Comparing these different sets of simulations provides important information about the energy deposition in the atmosphere as a function of the low energy cut-off E$_\textrm{C}$. 
We discuss the results from the runs with T$_{\textrm{LT}}$ = 1 MK and 3 MK separately in the following two sections.

\subsubsection{15Mm, 1MK loop-top temperature}
\label{Sect:3.1.1}
The low energy electrons (E$_{c}$ = 5 keV;  left panels of Fig.~\ref{Fig:atm_resp_15Mm_1MK}) mainly deposit their energy in-situ in the corona and TR, causing an increase of temperature (panel \emph{a}) and density (panel \emph{d}) in the corona up to around 1 order of magnitude in the first 10~s and the position of the TR to recede back towards higher column masses (or lower depths). The TR adjusts in column mass until the resulting increased radiative losses in the TR can balance the incoming thermal conductive flux from the coronal energy deposition.
The electrons lose all their energy and are completely stopped at heights between z~$\approx$~1.5--2 Mm. The temperature increase due to the beam heating causes an increase of pressure and therefore very large upward motions (up to~-500 km/s) of high temperature plasma (above 10$^{6}$ K), increasing with atmospheric height. At the same time, the pressure gradient drives a significant downflow of TR plasma ($\approx$~50 km s$^{-1}$) towards the chromosphere.  
After the heating is switched off, the atmosphere starts cooling off, initially from the lower regions where the denser plasma radiates more efficiently.

Electron distributions with E$_\textrm{C}$ = 10 and 15 keV (middle and right panels of Fig.\ref{Fig:atm_resp_15Mm_1MK}) contain more energetic electrons which are able to propagate through the corona without losing much energy and penetrate deeper into the atmosphere. As a result, there is no (for E$_\textrm{C}$ = 15 keV) or little (for E$_\textrm{C}$ = 10 keV) direct heating of the corona compared to the E$_\textrm{C}$ = 5 keV run, but most of the energy is directly deposited in the TR and chromosphere. This can be best seen in the bottom panels (\emph{m}--\emph{n}), showing the evolution of energy deposition per mass as a function of height. In the first few seconds, the beam energy deposition is maximum between 1.5 and 2 Mm. Hydrogen and helium ionize quickly and the radiative losses can no longer balance the large heating deposited there, resulting in an large increase of temperature localized in that region. 
The overpressure then drives an upflow of plasma from the low atmosphere towards the corona, similarly to the E$_\textrm{C}$ = 5 keV case discussed above.

We note that the upflow velocities are lower in the simulations with E$_\textrm{C}$ of 10 and 15 keV than the one with 5 keV.  The higher energy electrons can in fact penetrate deeper into the chromosphere where the plasma is denser. This means that plasma is harder to accelerate, but also that it will radiate the energy away more efficiently, resulting in a lower pressure increase. Both effects contribute to produce lower upflow velocities for the 10 and 15 keV  compared to the 5 keV case.  The upflows of plasma in the 10 keV and 15 keV simulations also causes the TR to be pushed out towards greater heights. 

In all three cases, the loop density increases over time because of the strong upflows of plasma towards the corona (\emph{chromospheric evaporation}), and consequently the loops are filled with high temperature plasma. As a result, the electrons start depositing more energy at increasingly higher atmospheric layers, as can be seen from the heating deposition rate plots (panels \emph{l}--\emph{n}). 

We note that the low energy electrons are more effective at heating the corona and driving evaporation of high temperature plasma, in agreement with what was found earlier by \citet{Peres87} and more recently by T14 and by \citet{Reep15} in their study of chromospheric evaporation as a function of different electron properties for flare-size events. 

\subsubsection{15Mm, 3MK loop-top temperature}
\label{Sect:3.1.2}
Figure~\ref{Fig:atm_resp_15Mm_3MK} shows the atmospheric response to the EB heating models for the 15 Mm half-loop with initial T$_{\textrm{LT}}$~=~3 MK.  As mentioned before and also shown in Fig.~\ref{Fig:init_atm}, the initial loop-top density is almost one order of magnitude higher than in the 1MK loop.
Because of the higher density,  more beam energy is deposited in the corona compared to the  T$_{\textrm{LT}}$~=~1 MK loop, regardless of the electron energy.  In the E$_\textrm{C}$ = 5 keV run (left panels), a significant fraction of beam energy is released in the corona (see panel \emph{l}), causing an increase of temperature and pressure driving an upflow and downflow of plasma away from the site of maximum energy release ($\approx$~10 Mm) in the first $\approx$~15~s.  Not all the electrons are stopped in the corona, but some of them are able to reach the TR and deposit their energy directly at an atmospheric height of z~$\approx$~1.5 Mm.  At that location, the combination of energy deposition and a thermal conduction front from the corona causes a pressure gradient and resulting upflowing plasma towards the corona and a dowflowing plasma at low speeds in the TR from around 10~s onwards. 

In the E$_\textrm{C}$~=~10 and 15 keV simulations (middle and right panels), the beam energy is initially deposited in the TR (as shown in panels \emph{m} and \emph{n}), causing upflows of plasma from that region towards the corona. As the loop starts being filled with high temperature plasma, the electrons get stopped at progressively greater heights, and as a consequence the evaporation front travels upward along the loop (see panels \emph{h}--\emph{i}).

The results above indicate that, for the same heating model, the flows and density/temperature increase in the lower atmosphere are much lower for a denser and hotter loop, where the energy is dissipated more efficiently in the coronal part of the loop. In addition, the response of the atmosphere to the heating by low energy electrons or thermal conduction  (see below Sect.~\ref{Sect:3.2}) is much slower (after~$\sim$~10~s), because of the higher loop density, which means that the mass is harder to accelerate.
\begin{figure*}[!h]
\centering
\includegraphics[width=0.7\textwidth]{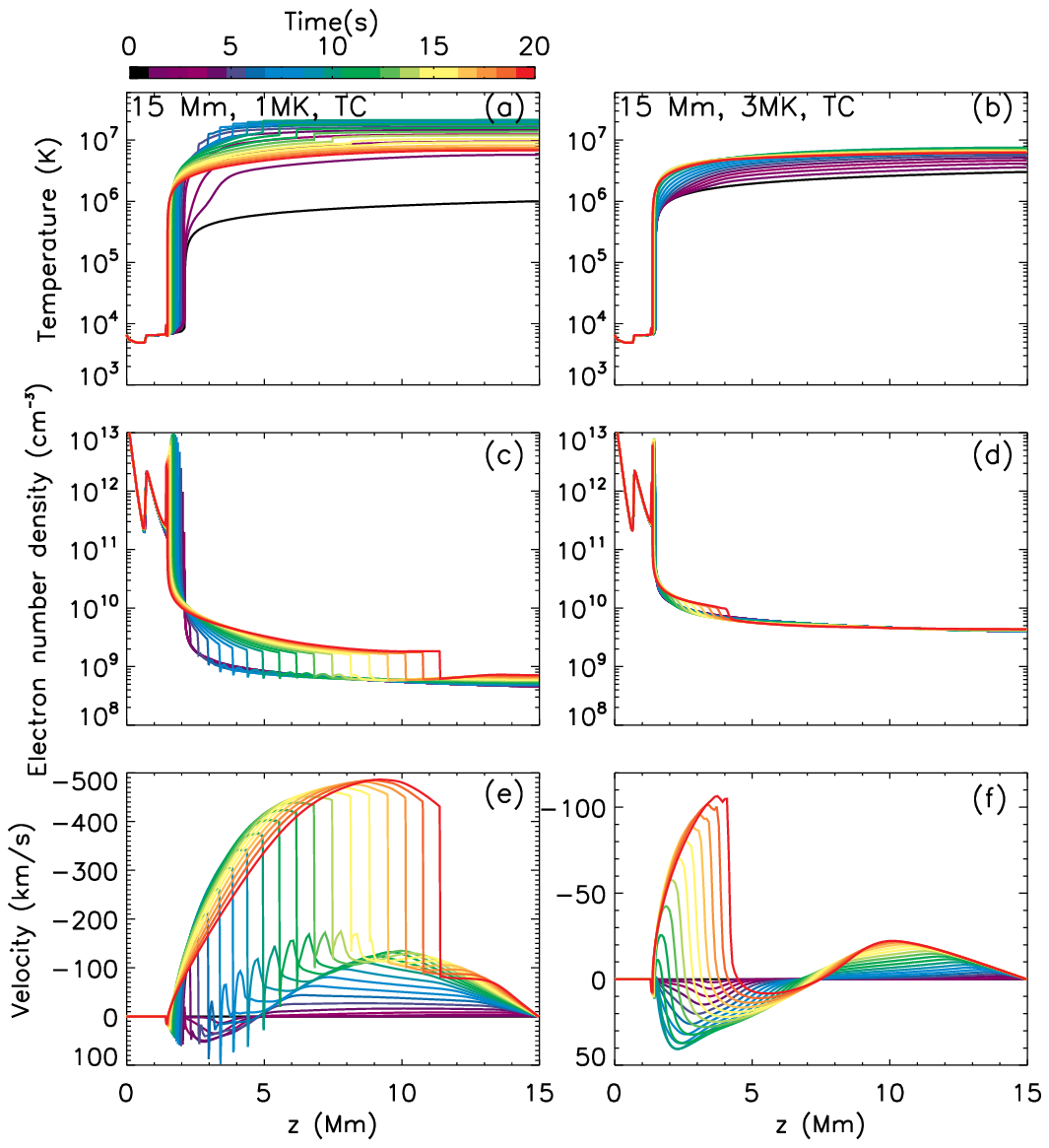}
\includegraphics[width=0.75\textwidth]{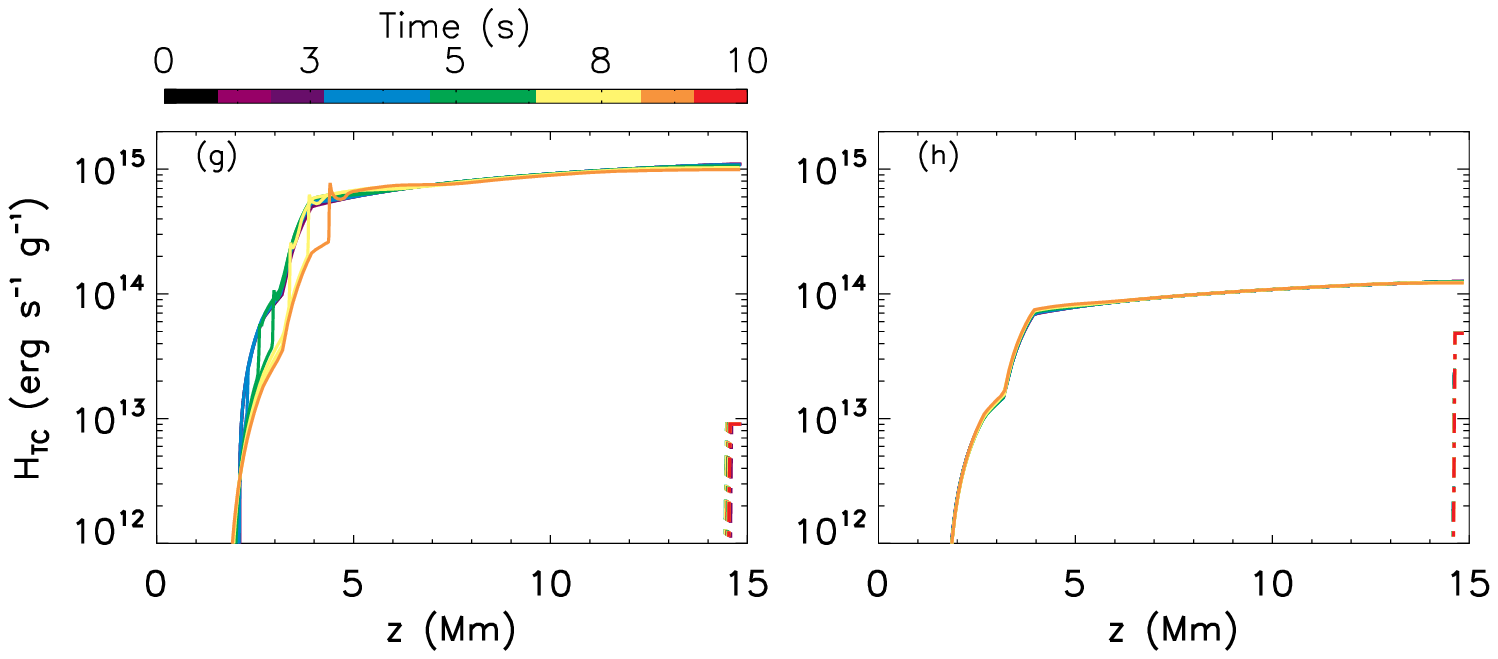}
\caption{Atmospheric response to the RADYN nanoflare thermal conduction simulations for the 15 Mm loop with initial loop top temperature of 1MK (left panels) and 3MK (right panels).  The different panels show (from top to bottom): the plasma temperature, electron density,  bulk velocity and heating rate per mass (\emph{H$_\textrm{TC}$}). The dotted lines in panels \emph{g} and \emph{h} show the background heating per mass applied to maintain the initial steady loop atmosphere at the chosen pre-nanoflare temperature.  }  
\label{Fig:atm_resp_15Mm_th}
\end{figure*}
\subsection{Thermal conduction heating}
\label{Sect:3.2}

In this section we describe the evolution of the atmospheric variables in the RADYN nanoflare simulations assuming in-situ heating of the corona and subsequent transport through thermal conduction, without the presence of accelerated electrons. As mentioned in Sect.~\ref{Sect:2}, the total energy deposited locally in the corona is the same as the energy released by the accelerated electrons for the EB simulations.

Figure~\ref{Fig:atm_resp_15Mm_th}  shows the results of the TC simulations for the 15 Mm half-loop with 1 MK (left panels) and 3~MK (right panels) initial loop top temperatures. The figure shows (from top to bottom): the evolution of the plasma temperature, density, bulk velocity and heating rate per mass (that we call \emph{H}$_\textrm{TC}$). 

\begin{figure*}
\centering
\includegraphics[width=\textwidth]{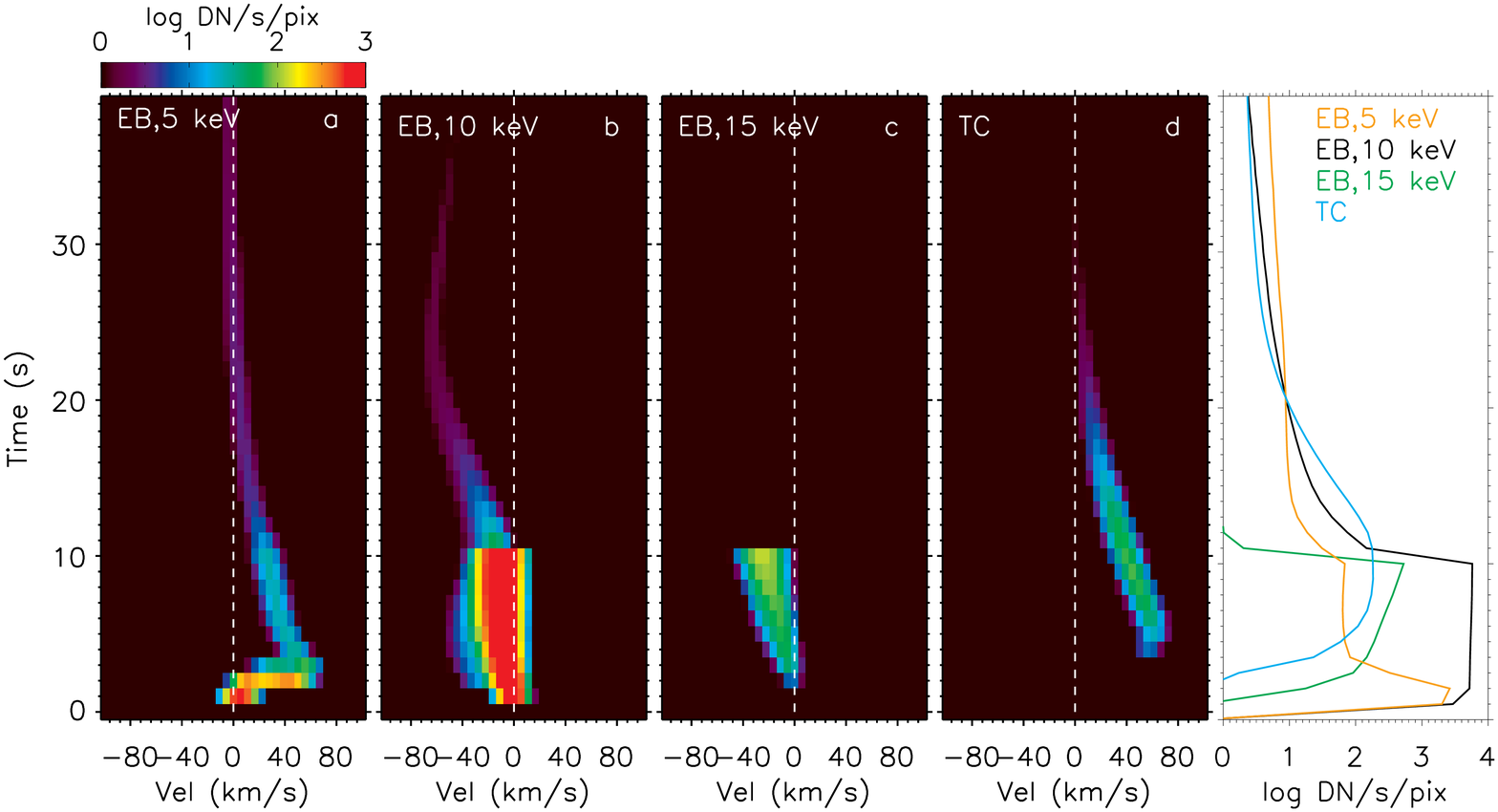}
\caption{Synthethic \siiv~spectra (a--d) and lightcurves (e) as a function of time as forward
modeled from the RADYN nanoflare simulations assuming 10s heating by: electron beams with E$_\textrm{C}$ of 5keV (a), 10keV
(b) and 15keV (c); and thermal conduction only (d) for the 15 Mm loop with initial loop top temperature of 1MK. Negative (positive) velocities indicate blue(red)-shifts}
\label{Fig:Si4_1MK}
\end{figure*}

We have explored the possibility of releasing the energy over different lengths in the corona. We note that, unless the heating is strongly concentrated at the foopoints \citep[e.g.][]{Mueller03,Mueller04,Testa05}, the details of the spatial distribution of the heating are not crucially affecting the plasma evolution, because the thermal conduction is efficient in re-distributing the energy in the corona. We show here the results for a simulation in which heating is distributed over the uppermost $\approx$~9~Mm of the loop.
In the cooler and less dense T$_{\textrm{LT}}$~=~1~MK loop, the heat causes a quick  increase of the coronal  temperature to about 20~MK in the first few seconds into the simulation. The conduction front then reaches and starts heating the TR after $\approx$~3~s, causing an increase of pressure driving very large upflows ($\sim$~-500~km$\cdot$s$^{-1}$) of hot plasma towards the corona (panel \emph{e}).  Figure~\ref{Fig:atm_resp_15Mm_th} (e) also shows systematic positive velocities in the TR  i.e., a downflow, which corresponds to redshifted TR lines (e.g., \siiv~lines observed by \emph{IRIS}, see Fig.~\ref{Fig:Si4_1MK}), as also discussed by T14.

In the T$_{\textrm{LT}}$ = 3~MK loop, the corona has a larger heat capacity and reaches a lower maximum temperature of $\approx$~7.5~MK. The heating is less effective and plasma flows are delayed, slower and less significant here. Moreover, as soon as the heating ends (t~$\sim~$10~s) the denser plasma abruptly begins to cool down and a pressure dip forms in the low corona, driving an initial downflow \citep{Reale16}, which then turns into an upflow when the chromospheric plasma takes over.

The atmospheric response to the heating in the TC models shows very similar results to the simulations with EB heating and E$_\textrm{C}$ = 5 keV discussed in Sect.~\ref{Sect:3.1}. This is also the case for the forward modeling of the plasma emission, as will be shown in Sect.~\ref{Sect:4}.

\section{Forward modeling of the optically thin and chromospheric emission}
\label{Sect:4}
We forward model the predicted emission from our simulations in the \emph{IRIS} \siiv~spectral line and in the  94~\AA~filter of AIA. The \siiv~line (T~$\approx$10$^{4.9}$K) is the brightest optically thin TR line observed by \emph{IRIS} and provides a powerful diagnostic of moss emission at the footpoints of the nanoflare loops. On the other hand, the AIA 94~\AA~channel is dominated, in the core of active regions, by plasma at around $\gtrsim$~4~MK from hot coronal loops \citep[e.g.][]{Testa12}.

The synthetic emission from the two instruments was calculated using the following formula:

\begin{equation}
\begin{split}
I\left(\lambda,T(z,t),N_{\textrm{e}}(z,t) \right)\ =G\left(\lambda,T(z,t),N_{\textrm{e}} (z,t)\right)\cdot \\
R_{\textrm{I}} \cdot N_{\textrm{e}}(z,t) \cdot N_{\textrm{H}} (z,t)\cdot dz(z,t),
\end{split}
\label{eq:1}
\end{equation}
where 
I$\left(\lambda,T(z,t),N_{\textrm{e}}(z,t) \right)$ is the intensity of the lines expressed in DN s$^{-1}$ pix$^{-1}$ as a function of the position \emph{z} along the loop and time \emph{t}. G$\left(\lambda,T(z,t),N_{\textrm{e}}(z,t)\right)$ represents the contribution function (ph cm$^{3}$ sr$^{-1}$s$^{-1}$) calculated using atomic data from CHIANTI v.8 \citep{Dere97, DelZanna15}, assuming coronal abundances \citep{Feldman92}, and standard \emph{SolarSoft} \citep{Freeland98} CHIANTI routines: \emph{emiss\_calc} for the forward modelling of the \siiv~line and \emph{isothermal} for the AIA~94~\AA, following the method described in the appendix of \cite{DelZanna11}. R$_{\textrm{I}}$ is the instrument response function (units: DN ph$^{-1}$ sr cm$^{2}$pix$^{-1}$), given by the product of the effective area, plate scale and gain of the telescopes. For AIA, we use a platescale of 8.46~$\cdot$~10$^{-2}$~sr pix$^{-1}$ while the effective area and gain were obtained using the \emph{aia\_get\_response(/dn)} routine. For \emph{IRIS}, we use 4 DN ph$^{-1}$  for the FUV spectrograph channel and the effective areas given by the \emph{iris\_get\_response} routine. 
Finally, N$_{\textrm{e}}$, N$_{\textrm{H}}$ and \emph{dz} represent the electron and hydrogen number density and grid size respectively, which are given by the RADYN models and vary as a function of  \emph{z} and \emph{t}. 
\begin{figure*}
\centering
\includegraphics[width=0.8\textwidth]{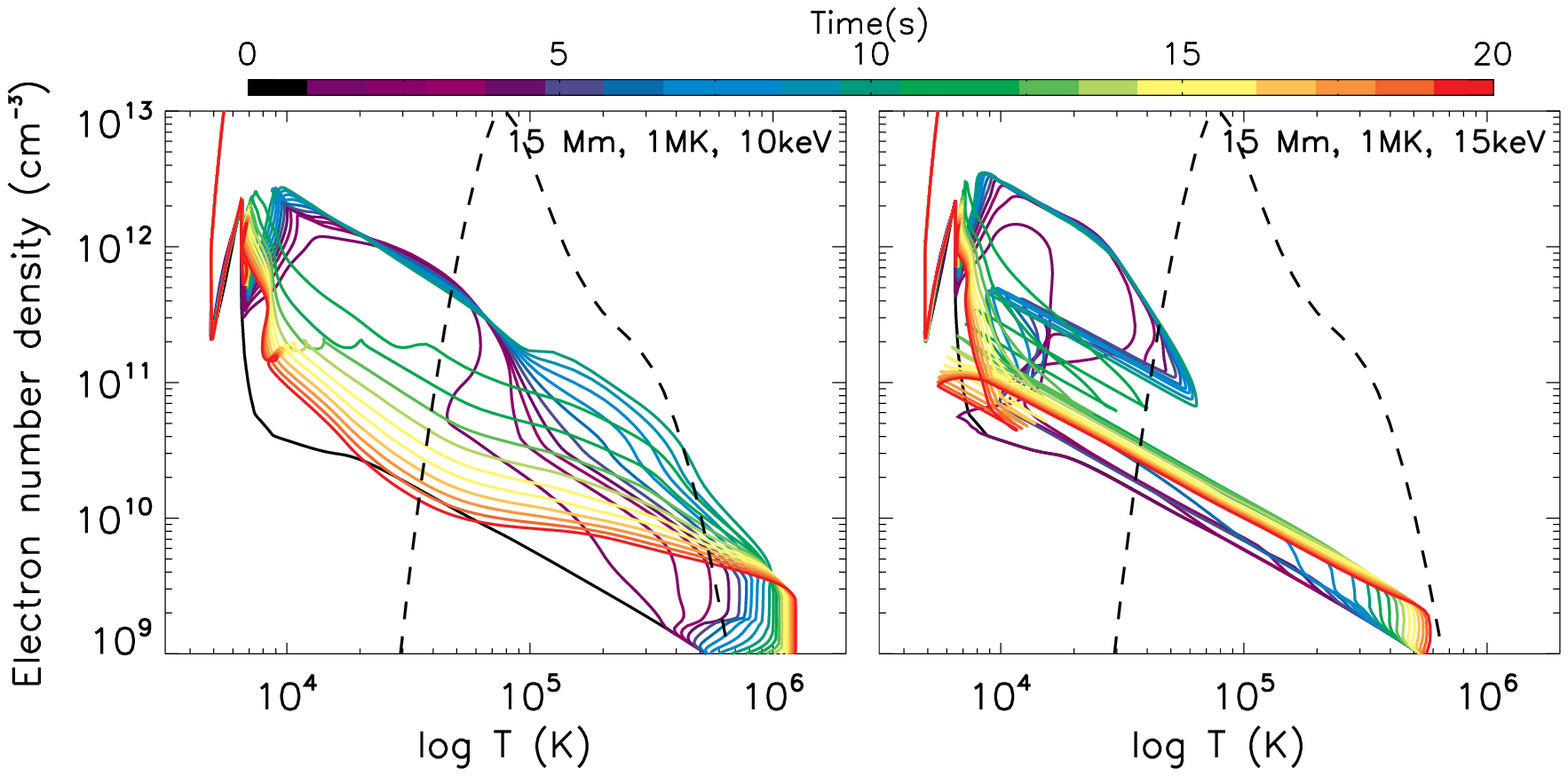}
\caption{Electron number density as a function of temperature for the simulations with the 15 Mm loop, apex temperature of 1~MK and EB model with E$_\textrm{C}$~=~10 (left) and 15~(right) keV. The contribution function of the IRIS \siiv~line is overplot in logarithmic scale. The plots show that right after 10~s (when the heating is switched off), the density drops significantly within the temperature range of formation of the ion.
}
\label{Fig:ne_T}
\end{figure*}

\subsection{IRIS \siiv~emission}
\label{Sect:4.1}
\begin{figure*}
\centering
\includegraphics[width=\textwidth]{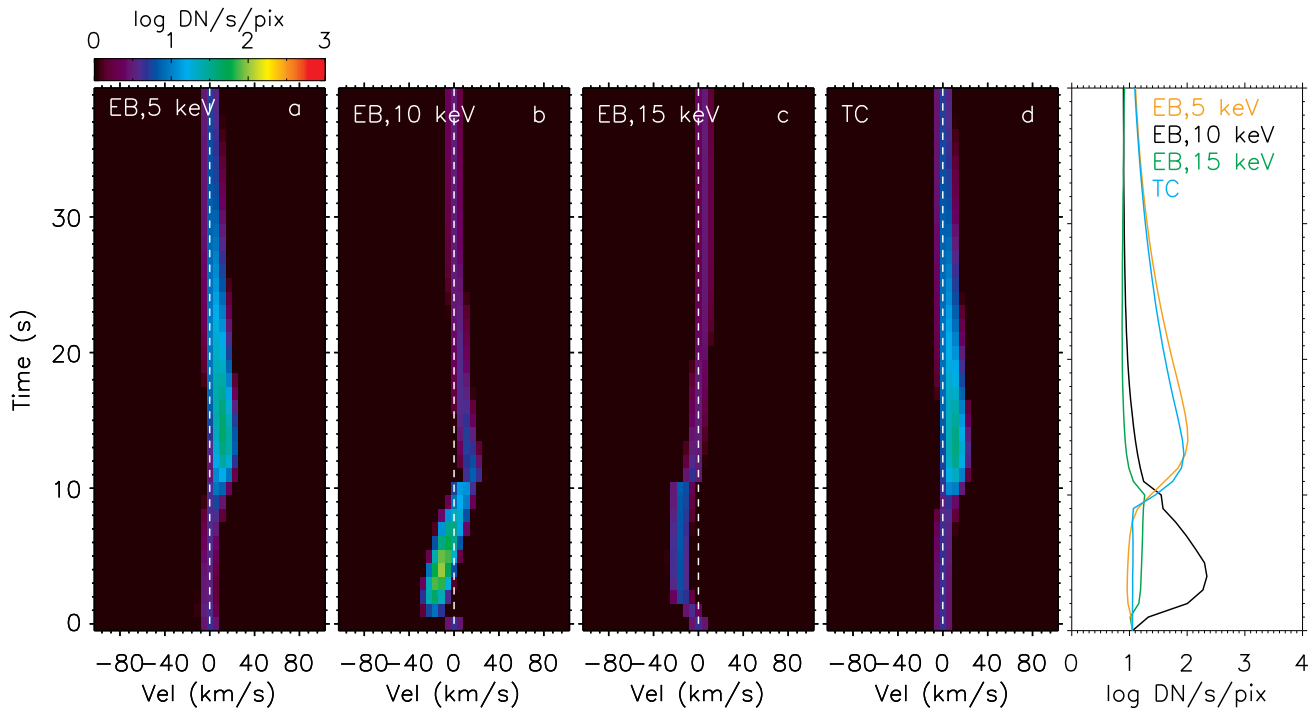}
\caption{Synthethic \siiv~spectra and lightcurves for the 15Mm loop with initial loop top temperature of 3MK.  Negative (positive) velocities indicate blue(red)-shifts. See caption of Fig.~\ref{Fig:Si4_1MK} for more details. }
\label{Fig:Si4_3MK}
\end{figure*}

Using Eq.~\ref{eq:1}, we have synthesized the evolution of the \siiv~spectra for our different nanoflare models. We assume that the loop is observed from above and thus the \siiv~emission from the  loop footpoints is spatially integrated along the line of sight and is observed in one \emph{IRIS} pixel. The synthetic spectra are obtained including an instrumental width for the FUV channel of 31.8~m\AA\footnote{iris.lmsal.com/itn26/codes.html}. Figures~\ref{Fig:Si4_1MK} and \ref{Fig:Si4_3MK} show the evolution of the \siiv~spectra as a function of time (Y-axis) and Doppler shift velocity (X-axis) for the first 40 s into the nanoflare simulations. Negative (positive) velocity indicates blueshifts (redshifts), consistent with Figs.~\ref{Fig:atm_resp_15Mm_1MK}, \ref{Fig:atm_resp_15Mm_3MK} and \ref{Fig:atm_resp_15Mm_th}. The intensity of the line (in logarithmic scale) is shown in reversed colors as indicated by the colorbar on the top of panel \emph{a}. From left to right, the first 4 panels (\emph{a}-\emph{d}) present the \siiv~synthetic spectra for the EB simulations with E$_\textrm{C}$=5, 10 and 15 keV, and the TC simulation respectively. The last panel \emph{e} shows the \siiv~light curves, obtained by integrating the emission over the spectral line, for the four simulations listed above.  

In Fig.~\ref{Fig:Si4_1MK}, both the intensity images and the light curves show a sudden increase of intensity of the \siiv~line over the first few seconds in response to the heating. The \siiv~width also increases by a factor of two or more (differently for different heating models) as a result of the superposition of different line components (shifted and at rest)  along the line of sight. In particular, the intensity rise is more dramatic in the EB simulations, where the emission increases up to 3--4 orders of magnitude over the first 10 seconds, before dropping dramatically to its pre-heating value. In order to better understand this sudden drop in intensity, in Fig.~\ref{Fig:ne_T} we show the electron number density as a function of temperature for the simulations with the 15 Mm loop, apex temperature of 1~MK and EB model with E$_\textrm{C}$~=~10 (left panel) and 15~(right panel) keV.  The dotted curve overlaid on the figure represents the contribution function of the \siiv~line, which has been normalized to arbitrary units. This figure shows that after 10 s (when the heating is switched off), the electron density drops significantly within the temperature range of formation of the \siiv~line, justifying the rapid decrease in intensity in the spectra in Fig.~\ref{Fig:Si4_1MK}. This sudden variation in the \siiv~intensity is compatible with the observations of short-lived (10--30 s) TR brightenings in active region heating events \citep[e.g.][T14, Testa et al. in prep]{Testa13}, as discussed in the introduction. In addition, the maximum \siiv~intensity increase and line broadening are caused with the 10 keV electrons, which deposit a large fraction of their energy at a depth where most of the \siiv~emission is formed. We note that both the line intensity and width (in addition to the Doppler shift) represent crucial parameters which can help distinguish between different heating models. 

In the E$_\textrm{C}$~=~5 keV simulation (panel \emph{a}), we observe a steep increase of intensity and a large \siiv~redshift up to around 50 km$\cdot$s$^{-1}$ in the first $\approx$~3s. As explained in Sect. \ref{Sect:3.1.1}, at this time the electrons deposit a significant fraction of their energy in the TR, causing a large downflow of plasma as a consequence of the pressure gradient. A large \siiv~redshift is also observed in the TC simulation (panel \emph{d}) after around $\approx$~3s, in agreement with what we described in Sect. \ref{Sect:3.2}.

In the 3~MK loop simulations (Fig. \ref{Fig:Si4_3MK}), the intensity increase of the \siiv~line is much lower than in the 1~MK case described above (note that we used the same intensity color scale for both Figs.~\ref{Fig:Si4_1MK} and \ref{Fig:Si4_3MK}). The 3~MK loop is in fact significantly denser than the 1MK loop and as a result, the electron energy is dissipated more efficiently before reaching the TR and chromosphere, as also discussed in Sect. \ref{Sect:3.1.2}. The EB simulations with  E$_\textrm{C}$~=~10 and 15 keV (panels \emph{b} and \emph{c}) show a moderate blueshift (up to 30 km$\cdot$s$^{-1}$) of the \siiv~line profile in the first 10 s, while in the E$_\textrm{C}$=5 keV and TC cases (panels \emph{a} and \emph{d})  the line appears to be slightly redshifted ($\approx$ 20 km$\cdot$s$^{-1}$) after 10 s. As discussed in Sects.~\ref{Sect:3.1.2} and \ref{Sect:3.2}, in these cases the TR is in fact heated by the thermal conduction front from the corona only after $\approx$~10~s into the simulations. 

Figures~\ref{Fig:Si4_1MK} and \ref{Fig:Si4_3MK} show that the evolution of the \siiv~spectra is very sensitive to both the details of the heating model and initial atmosphere, as also shown and discussed in T14. The results for the other nanoflare models with half-loop lengths of 50 and 100 Mm,  T$_{\textrm{LT}}$~=~5MK and different total energy will be discussed in Sect. \ref{Sect:6}.

\subsection{Coronal emission}
\label{Sect:4.2}
In contrast to the TR footpoint emission, the modeling of the coronal emission from the nanoflare loops observed by AIA requires some additional considerations: 

(1) To take into account the fact that the loop cross-section (\emph{A}~=~5$\cdot$10$^{14}$ cm$^{2}$) is much smaller than the AIA pixel area (\emph{A$_{AIA\_pix}$}$\approx$~1.89$\cdot$10$^{15}$ cm$^{2}$), we multiply Eq.~\ref{eq:1} by an additional factor, given by the ratio of these two values (\emph{A}/\emph{A$_{AIA\_pix}$}~$\approx$~0.26).   (2) We note that the AIA 94~\AA~passband is sensitive to both hot emission (from \fexviii~($\approx$~7~MK) and cooler emission \citep[e.g.][]{Boerner14,Testa12b,DelZanna12,Martinez11,ODwyer10}. 
To isolate the emission from the hot core AR loops, we only model the emission above 3~MK due to \fexviii. This approach allows us to directly compare the results of our simulations to both imaging observations with AIA, where the \fexviii~emission is often isolated by removing the cooler contaminating component from the 94~\AA~channel, and spectroscopic data from instruments such as \emph{Hinode}/EIS, observing emission lines formed at similar temperatures as the \fexviii~\citep[see for instance ][]{Ugarte14, DelZanna13, Testa12, Reale11}. (3) We calculate the total emission per AIA pixel by simply dividing the total emission integrated along the coronal part of the loop (where the temperature is above 3MK) with the corresponding number of AIA pixels along that length. While this assumption does not take into account any geometrical effects (which vary from case to case depending on the line of sight of the observations and on the inclination of the loops), it provides a lower limit of the emission that can be observed in one AIA pixel.

\begin{figure*}
\centering
\includegraphics[width=0.9\textwidth]{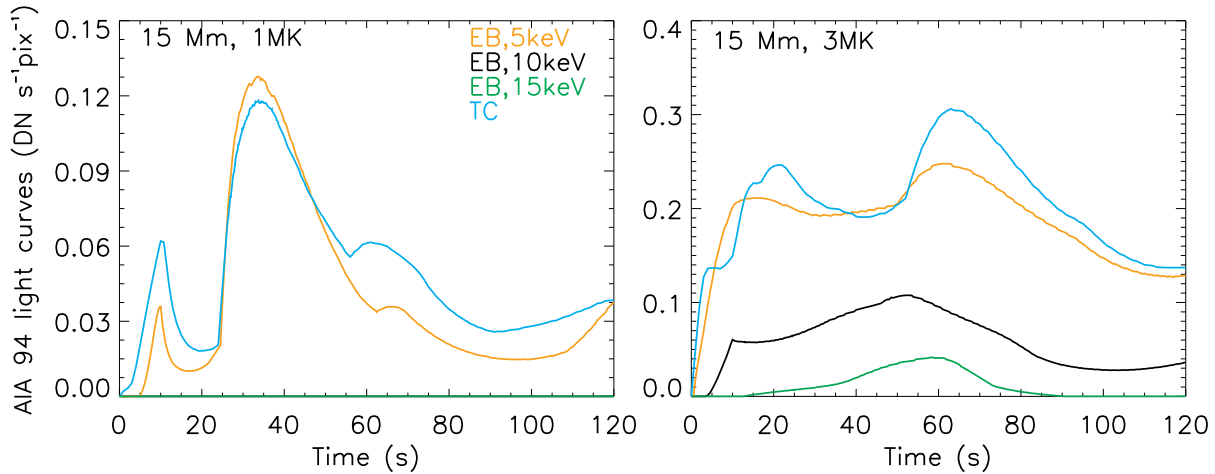}
\caption{AIA synthetic light curves in the 94~\AA~filter for our nanoflare models with 15 Mm  loop length and T$_{\textrm{LT}}$= 1MK loop (left) and T$_{\textrm{LT}}$ = 3MK (right), showing only plasma emission above 3MK. Different colors represent light curves for different heating models, as indicated by the legend on the left panel. The light curves for the models with loop apex temperature of 1~MK and E$_\textrm{C}$~=~10 and 15 keV in the left panel are identical to zero because in these simulations the coronal plasma temperature does not reach 3~MK.
}
\label{Fig:aia}
\end{figure*}

Figure~\ref{Fig:aia} shows the AIA synthetic light curves (DN~s$^{-1}$~pix$^{-1}$) in the 94~\AA~filter for the 15 Mm half-loop length with T$_{\textrm{LT}}$~=~1MK loop (left panel) and T$_{\textrm{LT}}$ = 3MK (right panel) and different heating models, for the first 120~s into the simulations.  The figure indicates that the light curves for the T$_{\textrm{LT}}$~=~3MK loop are higher (by a factor of at least 2) than those for the T$_{\textrm{LT}}$~=~1MK loop, regardless of the properties of the heating model, due to the significantly higher initial temperature and density of the initial loop. 

In addition, as discussed in the previous sections, the electrons deposit more energy in the corona in the case of the denser 3~MK loop than in the almost empty 1~MK loop. We also note that the light curves for the models with loop apex temperature of 1~MK and E$_\textrm{C}$~=~10 and 15 keV (left panel) are identical to zero because in these simulations the coronal plasma temperature does not reach 3~MK, which we have used as threshold value for calculating the 'hot' 94~\AA~emission.

 Figure~\ref{Fig:aia} also shows that, for both loop temperatures, the simulations with TC only and EB heating with E$_\textrm{C}$~=~5 keV produce higher 94~\AA~emission than the EB simulations with E$_\textrm{C}$~=~10 and 15 keV. As discussed in Sect.~\ref{Sect:3}, this is due to the fact that in the first two models most of the heating is deposited directly in the corona, causing a larger increase of the coronal temperature.
 
\begin{figure*}
\centering
\includegraphics[width=\textwidth]{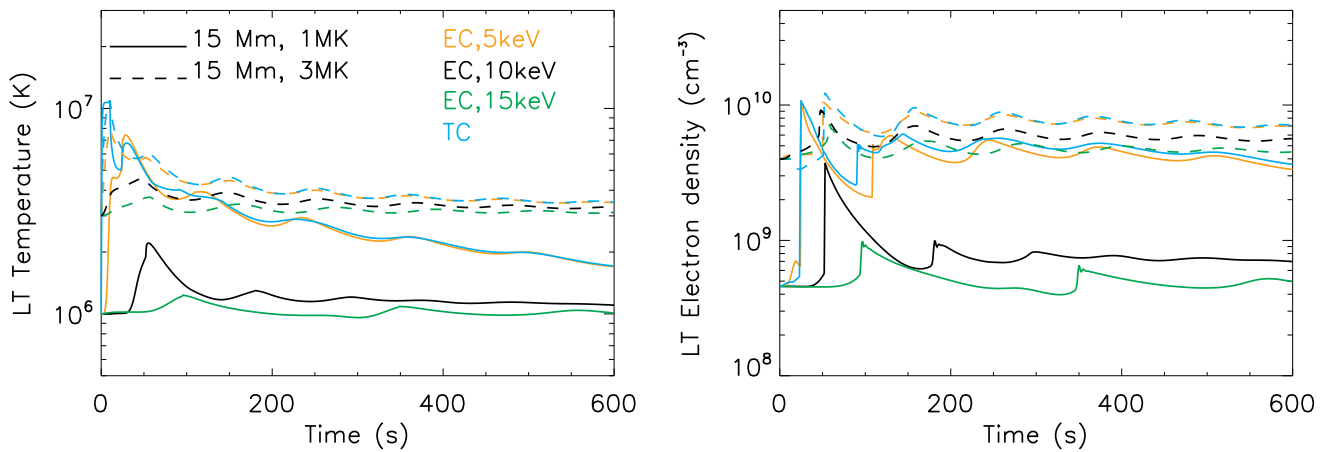}
\caption{Apex temperature (left) and electron number density (right) for the 15 Mm loop run as a function of time for the first 600~s into the simulations. Different colors represent different heating models, whereas different line styles (continuous and dashed) indicate different initial temperature for the loops, as indicated in the legend on the left panel. }
\label{Fig:tgtime}
\end{figure*}

The temporal evolution of the light curves show a first peak between 10--20~s, when the heating is deposited, and a second more intense peak between 40--60~s into the simulations following the chromospheric evaporation, due to loop being filled with hot upflowing plasma from the footpoints. After this time, the light curves first decrease (in the first 200~s) and then remain constant over time, in a similar fashion as the temperature and density evolution at the loop top. This can be best seen in Fig.~\ref{Fig:tgtime}, showing the apex temperature (left) and density (right) as a function of time for the different models. The curves show an oscillating pattern, which is the result of the evaporating plasma being reflected at the upper boundary of the loop, mimicking the effect of a symmetrical loop footpoint.

The AIA light curves in Fig.~\ref{Fig:aia} show that the peak intensity for a 15 Mm loop is observed to be within 100~s after the heating has taken place, while in observations the peak of the AIA 94~\AA~emission for the hot loops is often observed on longer time scales (from few to tens of minutes; Testa et al. in prep.). This is  not surprising and could indicate that multiple impulsive heating events over time might be needed to reproduce the observed light curves, i.e. a "nanoflare train" \citep[e.g.][]{Klimchuk06,Bradshaw12,Reep13a}. In addition, longer loops also produce AIA emission for a longer time; for instance, in our simulations with 100 Mm and apex temperature of 3~MK, the peak intensity in the 94~\AA~filter occurs around 400~s, as discussed in Sect.~\ref{Sect:6.1}.

The time delay between the TR and coronal emission can in principle provide important information about the mode of energy transport and the loop conditions. For instance, for a denser loop and 5~keV and TC simulations, the delay between the peak of the TR and coronal emission is smaller than in the case of a less dense loop, as can be seen by comparing Figs.~\ref{Fig:Si4_1MK} and \ref{Fig:aia}. However, in some cases such delays may be too short to be observed by AIA, which has a temporal cadence of $\approx$~12~s.

Further, we point out that the count rates shown in Fig.~\ref{Fig:aia}  for our single nanoflare loop models, although representing a lower limit, are a couple of orders of magnitude lower than the typically observed values $\sim$~20--30 DN s$^{-1}$ pix$^{-1}$, as observed for the coronal loops overlying the footpoints brightening observed by \citet{Testa13} and T14 \citep[see also][]{Testa12,Ugarte14}. We can thus provide a rough estimate of approximatively 100--200 loop strands which would be needed for reproducing the typical AR core observations of the coronal plasma. This estimate is consistent with previous calculations of e.g. \cite{Peter13}. This might also explain the delayed observed intensity peak: the real light curve would be an envelope of pulsed light curves slightly time-shifted one from the other as the heat pulses are likely not synchronous from one strand to the other \citep[e.g.][]{Reale12, Tajfirouze16}. We also note that the events showing moss variability associated with brightening of hot coronal loops typically involve a large number of footpoints. Many loops appear to be heated at slightly different times, and while the footpoint brightenings are well separated due to their short duration, the coronal emission has significantly longer timescales and many loop strands can contribute to the coronal emission observed in a pixel at a given time. For instance, for the events discussed in T14 and \citet{Testa13}, the footpoint brightenings cover an overall area of~$\approx$~10$^{16}$--10$^{17}$~cm$^{2}$ (i.e., up to several hundred times our assumed loop cross-section). Therefore, the overall observed hot emission (peaking around ~25 DN~s$^{-1}$pix$^{-1}$ in the 94~\AA~ AIA band) is compatible with the values  predicted by the simulations (Fig.~\ref{Fig:aia}). 

Furthermore, higher total energies for the heating events could also produce higher coronal emission (see Sect.~\ref{Sect:6.3}). Finally, a likely scenario is also that of a hybrid model in which a portion of the energy of the heating episode is released locally in the corona, even in events in which significant non-thermal particles are present (and might dominate the chromospheric and TR heating). Another possible explanation for the low coronal emission in our single nanoflare models might be related to the choice of uniform cross-section for the loops in our simulations. \cite{Mikic13} showed in fact that loops with non-constant cross-sections are more likely to develop thermal nonequilibrium, resulting in a significantly enhanced coronal emission. 

The fact that several loop strands (of the order of 100--200) might be needed to reproduce the AIA observations suggests that the cumulative hard X-ray emission emitted by the accelerated electrons in those strands might be strong enough to be detected by current instruments. T14 (Supplementary text S3) calculated the predicted X-ray emission observed by \emph{RHESSI} for a single nanoflare loop (with the same total energy as in our work) to be around 0.1 count for 10 s integration time and per detector. The nanoflare events are likely not occuring simultaneously, and considering that 100--200 events might occur during an integration time of 30 s, that would result in about 30--60 total counts/detector or 1--2 counts/s/detector. Therefore we expect most of these events to be below or close to the \emph{RHESSI} sensitivity limit \citep[of the order of 2--7 counts/s/detector for a 30 s integration time, ][]{Saint-Hilaire09}, though some might be detected by \emph{RHESSI}. The \emph{The Nuclear Spectroscopic Telescope Array} \citep[\emph{Nu}STAR; ][]{Harrison13},  has a better sensitivity and lower background than \emph{RHESSI} and might be able to observe more of these nanoflare-size events.

\begin{figure*}
\centering
\includegraphics[width=\textwidth]{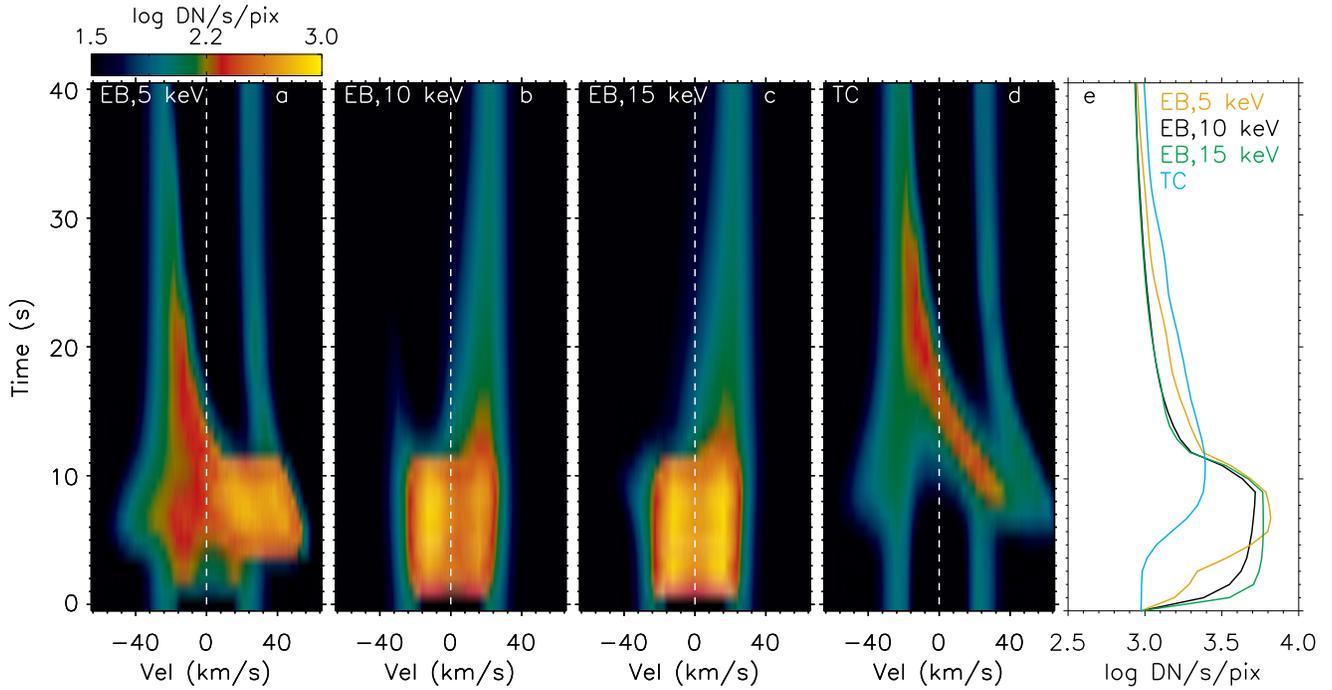}
\caption{Synthethic \mgii~h spectra for the 15Mm loop, initial apex temperature of 1MK and different heating models (see the caption of Fig.~\ref{Fig:Si4_1MK}).  The spectra are synthesized using the RH code with input atmosphere from the RADYN simulations, as explained in Sect.~\ref{Sect:5}. Negative (positive) velocities indicate blue(red)-shifts.   }
\label{Fig:Mg_1MK}
\end{figure*}

\begin{figure*}
\centering
\includegraphics[width=\textwidth]{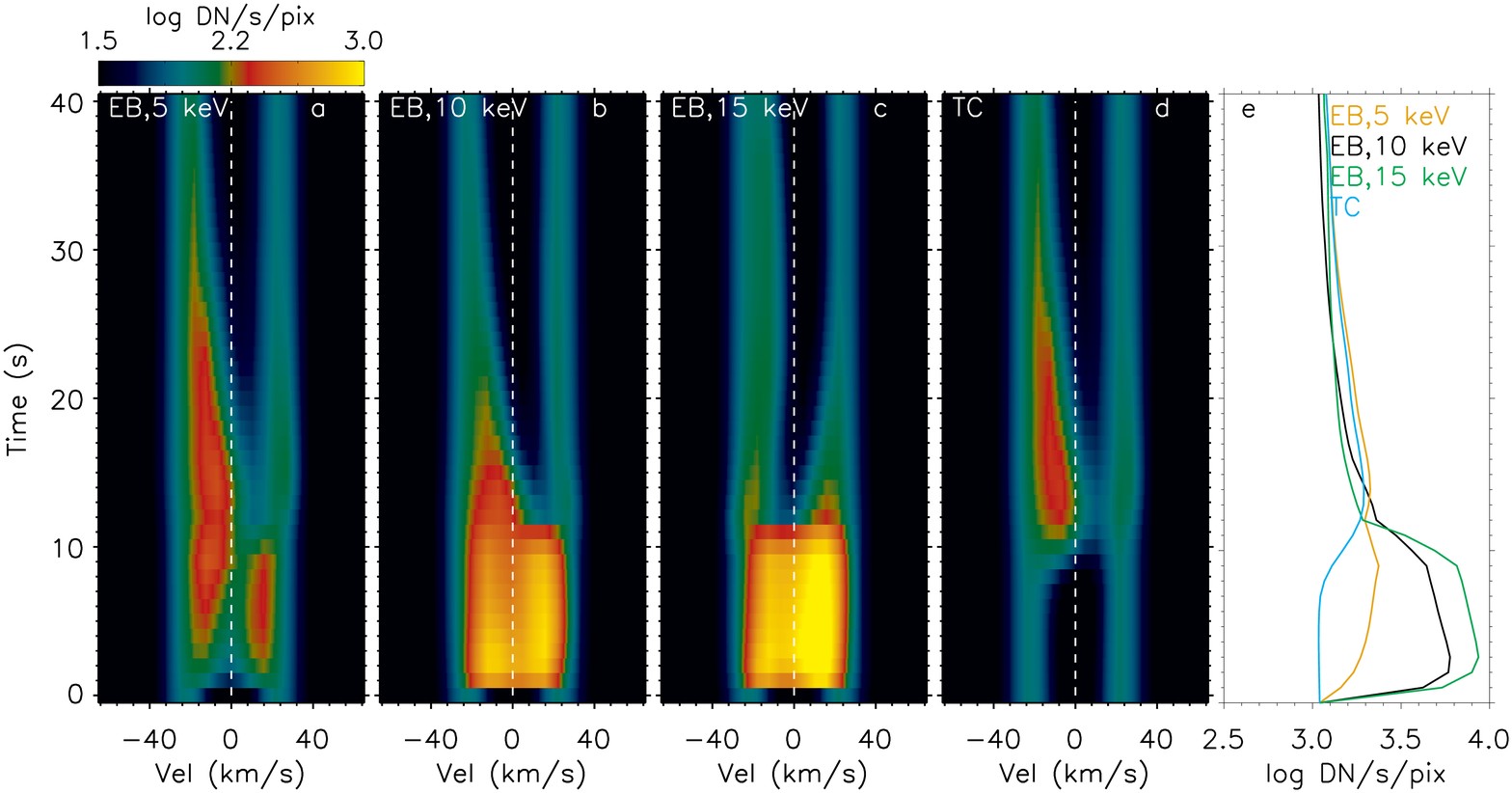}
\caption{Synthethic \mgii~h spectra for the 15Mm loop, initial apex temperature of 3~MK and different heating models.  See the caption of Fig.~\ref{Fig:Mg_1MK} for more details.}
\label{Fig:Mg_3MK}
\end{figure*}

\section{IRIS \mgii~chromospheric emission}
\label{Sect:5}
We synthesized the spectra of the chromospheric~\mgii~h~$\&$~k lines at  2803.53~\AA~and 2796.35~\AA~observed by \emph{IRIS} using the RH1.5D radiative transfer code \citep{Pereira15,Uitenbroek2001}. The RH code calculates the \mgii~lines in partial frequency redistribution (PRD). In PRD, the frequencies of the radiation absorbed and re-emitted by an atom are correlated.  We use a 21 level plus continuum model for the \mgii~atom. We note that the RADYN calculations do not include PRD, but rather assume complete frequency redistribution (CRD). Under the CRD assumption, the radiation scattering is completely not-coherent. While this is not appropriate for line synthesis (for which we use RH including PRD), as shown by \cite{Leenaarts13}, it is a reasonable approximation for the energetics in the RADYN simulations. 

We use snapshots of the atmosphere evolution calculated by RADYN every 1~s as input to the RH1.5D code for the first 40~s into the RADYN simulations. 
The output  spectra from the RH calculations (in units of J s$^{-1}$ m$^{-2}$ Hz$^{-1}$ sr$^{-1}$) were spectrally convolved to the \emph{IRIS} instrumental Gaussian profile 
 \citep[50.54~m\AA~in the NUV passband;][]{DePontieu14}, regrid into the \emph{IRIS} spectral sampling and converted into units of DN s$^{-1}$ pix$^{-1}$, using a gain of 18 DN ph$^{-1}$ for the NUV channel and effective area calculated as explained in Sect.~\ref{Sect:4}.

The \mgii~h~$\&$~k lines are formed over a large range of atmospheric heights and the interpretation of their spectra is not straightforward \citep[e.g.][]{Leenaarts13,Leenarts13b, Pereira13,Kerr15,Kerr16}. In the quiet sun, they are usually observed as double peaked with a pronounced central reversal core. The modeling of these lines is also challenging and the current models encounter difficulties in reproducing their observations in the AR plage or during flares \citep[e.g.][]{Carlsson15,Rubio17}.  
Here we aim to analyze the evolution of basic diagnostics of intensity and velocity as a function of our different nanoflare models which can be directly compared with the \emph{IRIS} observations, and postpone a more accurate modeling of the chromospheric emission and thorough analysis of the \mgii~spectral features to a following work. 

Figures~\ref{Fig:Mg_1MK} and \ref{Fig:Mg_3MK} summarize the results of synthetic spectra over time for the \mgii~h~2803.529~\AA~line, as a function of different heating models for the simulations with 15Mm half-loop length and initial apex temperatures of 1 and 3~MK respectively. Similar results are found for the \mgii~k~line and are thus not reported here. Different panels represent different nanoflare heating models, similarly to the \siiv~spectra images shown in Sect.~\ref{Sect:4.1}. 

The \mgii~h~line profile shows very peculiar features in the simulations with low energy cut-off and TC for the loop with apex temperature of 1MK (panels a and d of Fig.~\ref{Fig:Mg_1MK}). For these runs, the spectra are in fact characterized by multiple-peak profiles with oppositely-directed high velocity Doppler shifts (up to $\approx$~$\pm$~40 km$\cdot$s$^{-1}$),  with the redshifted component being significantly more intense than the blueshifted counterpart. This redshifted component is the result of a strong downflow of cool plasma observed in the first 10~s of the simulation, which is caused by the large pressure gradient formed around z~$\approx$~1.7~Mm (see Sect.~\ref{Sect:3.1}). Such peculiar profiles and strong shifts have not been commonly seen in observations so far. If observed, they might represent an indication of heating by low energy electrons or thermal conduction, as suggested by our models. However,  it should be pointed out that in observations we are more likely observing a superposition of \mgii~line profiles from the loops as well as background chromospheric emission from the plage. In particular,  a strong chromospheric heating may take place in the same AR (independently of the nanoflare heating) and determine the dominant emission for this line. 

For simulations with high E$_\textrm{C}$ ($\geq$~10 keV), the \mgii~ profile is mostly double-peaked for the first 10~s. The two peaks have a similar intensity and the centrally reversed core is at rest. After $\approx$~10~s, the spectra are dominated by the red wing, and the line profile becomes almost single-peaked. 

 For all the EB simulations in Fig.~\ref{Fig:Mg_1MK} (panels a--c), we note an increase of intensity in the first 10~s as a quick response to the heating, similarly to what was observed in the \siiv~synthethic spectra (Sect.~\ref{Sect:4.1}). 
 

In the TC simulation, the line reaches its peak intensity just after 10~s (see light curves in Fig.~\ref{Fig:Mg_1MK} e), because in this case the chromosphere is heated by a thermal conduction front from the corona, which has a longer timescale than the electron beam heating. 

The synthethic spectra obtained for the denser 15 Mm loop with apex temperature of 3~MK (Fig.~\ref{Fig:Mg_3MK}) show a number of different features compared to the spectra in Fig.~\ref{Fig:Mg_1MK} for a loop with apex temperature of 1~MK. First of all, the \mgii~ profile is always observed to be double-peaked and  the increase of intensity of the line as a function of E$_{\textrm{C}}$ is different compared to an almost empty loop. Specifically, for high E$_\textrm{C}$ ($\geq$~10 keV), the intensity increase for a denser loop is larger or comparable to the case of a low density loop, whereas for the 5 keV and TC simulations no significant \mgii~brightening or dramatic change in the line profile is observed as a consequence of the nanoflare event. In these latter cases the chromosphere is not heated significantly, as most of the energy is dissipated in the corona and partly in the TR, before reaching the lower atmosphere. In addition, the line does not show strong shifts like those observed in the TC and EB, E$_{\textrm{C}}$= 5 keV simulations of Fig.~\ref{Fig:Mg_1MK}. These differences are the result of the 1~MK loop being significantly less dense than the 3~MK loops. In a denser loop, the low-energy electrons are stopped at a higher height in the atmosphere and more energetic electrons are necessary to reach the chromosphere. 

Finally, we note that the relative intensity increase of the \mgii~line compared to the pre-flare conditions due to the nanoflare heating  is not as dramatic as in the case of the \siiv~line. The relative difference in the \mgii~synthethic intensity between the T$_\textrm{LT}$~=~1 and 3MK loops is also not as large as that observed for the \siiv~synthethic spectra.

The results presented here show that the analysis of the \mgii~lines can provide additional and different diagnostics of nanoflare heating to those provided by the TR \siiv~spectra. For instance, the intensity increase of the \mgii~due to the nanoflares is similar or stronger for the higher density than low density loops, in contrast to what is observed for the \siiv~line. In addition, thermal conduction heating alone is not as efficient in increasing the \mgii~intensity as the electron beam heating for either the low or high density loops. This is not true for the TR emission, where both thermal conduction and electron beam heating can produce a significant increase in the \siiv~line intensity. 

In this work, we have analyzed the general behavior of the \mgii~lines to different nanoflare models in order to demonstrate their diagnostic capabilities. A thorough analysis of all their spectral features \citep[also including the accurate modeling of the initial pre-nanoflare plage emission, see e.g.][]{Carlsson15} is deferred to a future paper.  
\section{Discussion: the parameter space}
\label{Sect:6}
In the previous sections, we discussed the impact of different heating properties and initial conditions of the loops on the atmospheric response and forward modeling of observables for our nanoflare simulations. In particular, we focused on discussing the results of varying the non-thermal electron energy distributions and investigating the difference between heating by electron beams and in-situ heating of the corona, in loops with two different apex temperatures. In this section, we investigate in more detail the effect of modifying the initial physical conditions of the plasma, i.e. varying the length of the loops as well as their initial temperature. In addition, we briefly discuss the effect of assuming a different total energy release on the model predictions and observables. 
\subsection{Other loop lengths}
\label{Sect:6.1}
We performed the same set of simulations presented in Sect. \ref{Sect:3} for loops with \emph{L}/2~=~50 and 100~Mm, initial T$_{\textrm{LT}}$ = 1--3~MK and corresponding loop-top electron number densities of 10$^{8.0}$--10$^{9.0}$ (for the 50~Mm loop) and 10$^{7.1}$--10$^{8.6}$ cm$^{-3}$ (for the 100~Mm loop). We will discuss here only the main differences between the results of these runs and the ones presented in the previous sections.

\begin{figure*}
\centering
\includegraphics[width=\textwidth]{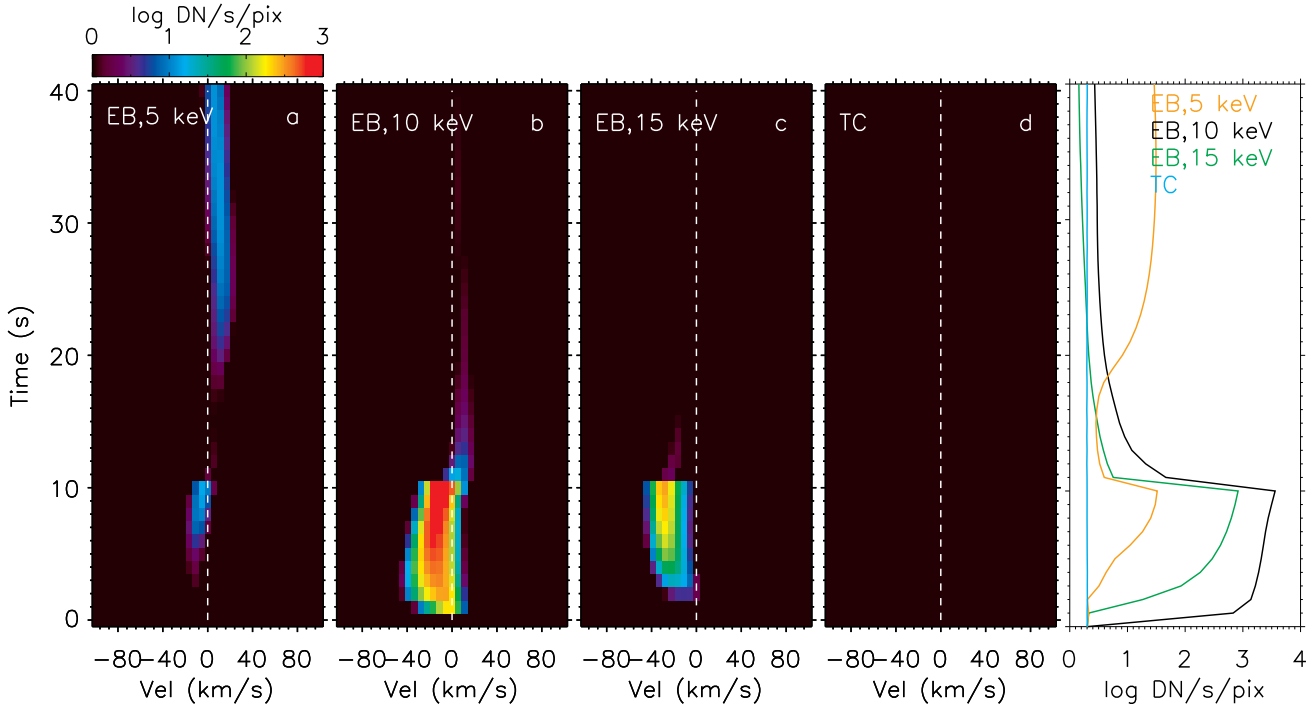}
\caption{Synthethic \siiv~spectra (a--d) and lightcurves (e) as a function of time for the 100 Mm loop with initial loop top temperature of 3~MK. Negative (positive) velocities indicate blue(red)-shifts. See caption of Fig.~\ref{Fig:Si4_1MK} for more details.}
\label{Fig:Si4_100}
\end{figure*}

The simulations with longer \emph{L}/2~=~50 and 100~Mm and T$_{\textrm{LT}}$~=~1MK show very similar results to the same runs with a shorter loop length of 15 Mm in the initial evolution of the nanoflare loop, whereas longer loops have a slower evolution on longer timescales (of the order of minutes). In particular, those runs present a similar behavior of atmospheric response as a function of low energy cut-off, i.e. the smaller the electron energy, the more energy deposited directly in the corona. Consequently, similar trends of blueshifts vs redshifts over time are observed, with the main difference being the Doppler velocity values, which become progressively larger for longer loops. This is a consequence of the plasma being less dense in the longer loops, resulting in larger flows. The similar overall trend is caused by the fact that when the loop is almost empty, the electrons stream through the corona almost collisionlessly. 

In contrast, the simulations with initial loop top temperature of 3 MK show more significant differences as a function of loop length. In particular, the intensity increase of the TR \siiv~line as a response to the heating is larger in the 50 Mm and 100 Mm half-loop length simulations, as shown in Fig.~\ref{Fig:Si4_100} for the 100~Mm loop. This is not surprising, as the lower density for these loops (compared to the 15 Mm loop with the same initial temperature) causes the electrons to have less collisions in the corona and deposit more energy into the lower atmosphere. The trend of Doppler shifts vs electron energy and heating model is similar to the 15 Mm case, with the only exception that redshifts in the \siiv~line are observed a bit later on ($\approx$~20~s) in the EB E$_\textrm{C}$ = 5 keV and TC simulations.  This is due to the fact that, for longer loops, it takes longer for the plasma downflow and the thermal conduction front from the corona to reach the lower atmosphere. In addition,  a small ($\approx$~20 km$\cdot$s$^{-1}$) but very faint blueshift of the \siiv~line is observed in the first 10~s of the E$_\textrm{C}$ = 5 keV run for the 50 Mm and 100 Mm loops (here we only show the \siiv~spectra for the 100~Mm loop as they are very similar). This initial blueshift is caused by the energy deposition of the electrons which are able to propagate through the corona and penetrate deeper down in the atmosphere. This is most likely not observed in the 15 Mm half loop simulations because in that case the plasma density is too high and a large part of the energy is dissipated in the corona without reaching the TR. Figure~\ref{Fig:Si4_100} also shows that the TC heating (panel d) does not affect the intensity of the \siiv~line. This is likely caused by the fact that the loop is very long (more than 6 times longer than the 15~Mm loop) and therefore most of the energy is dissipated along the coronal part of the loop without reaching the TR.

Another important difference between models with different loop lengths can be found in the synthetic AIA light curves, which differ both in their absolute values and time evolution. In particular, the 94~\AA~filter emission is at least one order of magnitude lower in the 50 Mm and 100 Mm  than in the 15 Mm half-loop length simulations. This decrease in intensity is directly due to the plasma density being orders of magnitude lower in the two longer loops.  In addition, the light curve peaks occur later on in these simulations because of the slower thermal conduction timescale and the longer time it takes for the evaporating hot plasma to fill the longer loops. 

Finally, although the details of the line profiles might change, the \mgii~spectra show overall similar results in simulations for loops of different length, i.e. similar light curves over time and peak velocities. This is because the details of the coronal part of the loop do not strongly affect the results of the dense chromosphere.
\subsection{Loop-top temperatures}
\label{Sect:6.2}
\begin{figure}
\centering
\includegraphics[width=0.45\textwidth]{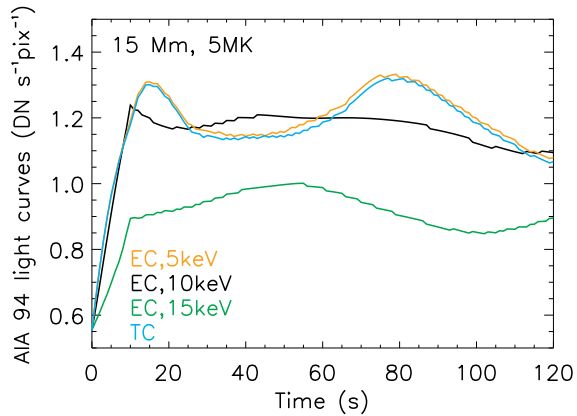}
\caption{AIA synthetic light curves in the 94~\AA~filter for our nanoflare models with 15 Mm  loop length and T$_{\textrm{LT}}$= 5 MK. See caption of Fig.~\ref{Fig:aia} for more details. }
\label{Fig:aia_5MK}
\end{figure}

We run a series of simulations with both electron beam and in-situ heating for 15 Mm half-length loops with initial apex temperature of 5~MK and electron number density of $\approx$~10$^{10.1}$~cm$^{-3}$. These runs aim to reproduce a scenario where heating is applied to a nanoflare loop strand which have been previously heated to very large temperatures. Similarly to the case of loops with T$_{\textrm{LT}}$ =~3~MK, the electrons are mostly stopped in the corona and very little energy can penetrate down to the TR and chromosphere. In addition, the in-situ local heating in the thermal conduction simulations is mainly dissipated in the corona.  As a result, no significant flows or intensity increase in the TR or chromospheric lines are observed. On the other hand, the synthethic AIA light curves are more intense (up to around one order of magnitude) than those obtained for the T$_{\textrm{LT}}$ = 1MK and 3 MK runs, as can be best seen in Fig.~\ref{Fig:aia_5MK}. The predicted increase of high temperature emission in the 94~\AA~filter is due to a combination of the initial coronal density being higher and the heating being mostly dissipated in the corona. 
\begin{figure*}
\centering
\includegraphics[width=\textwidth]{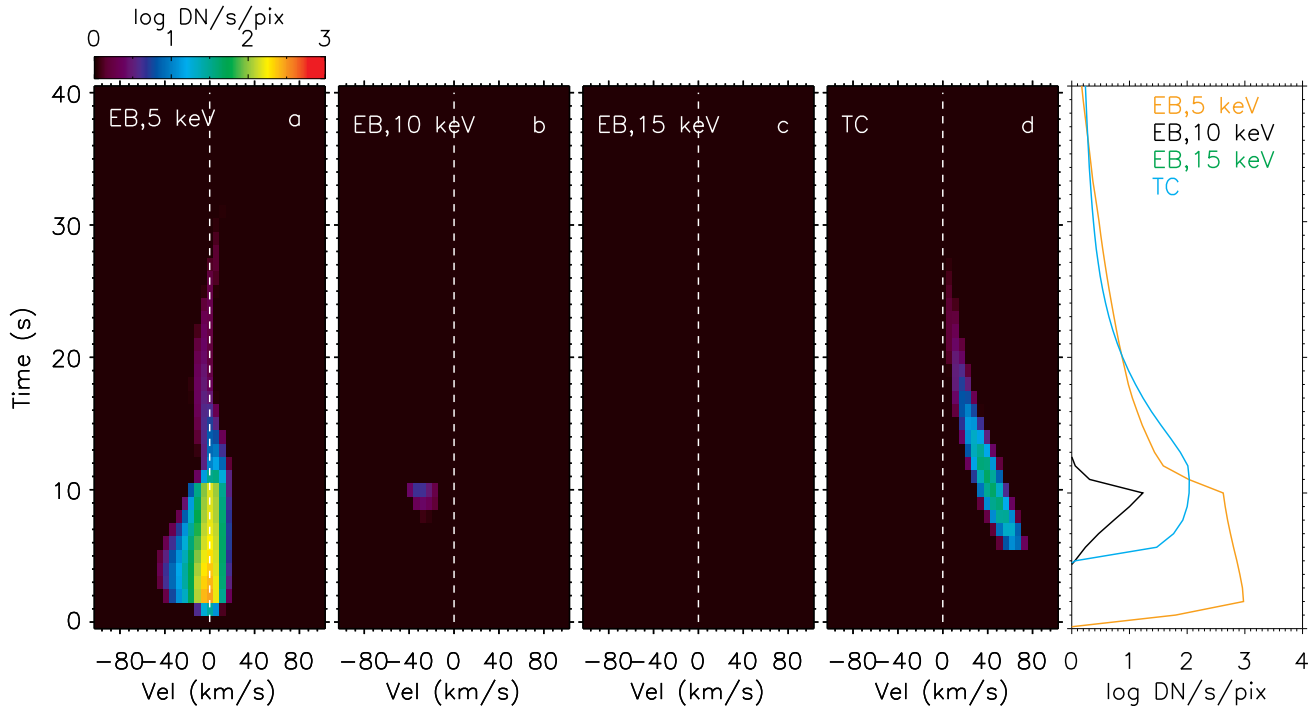}
\caption{Synthethic \siiv~spectra and lightcurves for the 15Mm loop with initial loop top temperature of 1MK and total energy release of E = 10$^{24}$ erg.  Negative (positive) velocities indicate blue(red)-shifts. See caption of Fig.~\ref{Fig:Si4_1MK} for more details. }
\label{Fig:sp_si4_1.e24}
\end{figure*}

\subsection{Varying the beam energy flux}
\label{Sect:6.3}
In order to investigate the importance of the nanoflare energy on the atmospheric response, we performed two additional sets of simulations assuming a total energy release of 10$^{24}$ and 10$^{25}$ erg for the loop with \emph{L}/2 = 15 Mm and initial apex temperature of 1~MK. These two values are representative of the range of energies that is typically assumed for nanoflare size heating events, and correspond (given the heating duration and loop cross-section described in Sect.~\ref{Sect:2}) to electron  beam energy fluxes $\emph{F}$ of 2~$\cdot$~10$^{8}$  and 2~$\cdot$~10$^{9}$ erg cm$^{-2}$ s$^{-1}$ respectively. Comparing electron distributions with the same spectral index and energy cut-off but different total energy fluxes corresponds to investigating the effect of varying the number of electrons in the beam at each energy. We also note that since the electron power-laws that we assume are quite steep ($\delta$ = 7), most of the electrons will have energy close to the cut-off value E$_\textrm{C}$, and therefore they will initially deposit their energy at the same atmospheric height, regardless of the total energy flux. 

The model runs that assume a total energy of 10$^{25}$ erg show essentially the same qualitative results as the simulations presented in  Sect.~\ref{Sect:3} for a nanoflare energy of 6~$\cdot$~10$^{24}$ erg. In particular, they both show the same trend of up/downflows at different atmospheric layers as a function of electron energy-cut off. 
The main difference between the two numerical investigations is that in the first case (E =10$^{25}$ erg) more energy is deposited in the atmosphere, causing faster plasma flows as well as a larger increase of density and emission measure for the optically thin plasma.

In contrast, the models assuming a smaller nanoflare energy (10$^{24}$ erg) show somewhat different and interesting results. As mentioned before, the electrons will initially deposit their energy mostly in the same atmospheric layer as in the simulations with a higher total energy flux. However, the energy deposited there will be significantly lower for the E = 10$^{24}$ erg than for the 6$\cdot$10$^{24}$ or 10$^{25}$ erg simulations. In the first case, the heating deposited is small enough that it can be more easily radiated away, avoiding the formation of a large overpressure region and the resulting strong increase of TR emission observed in the other two cases. In particular, the \siiv~intensity for the 10$^{24}$ erg simulation will be significantly less intense than in the 6$\cdot$10$^{24}$ erg case. Another important difference between models with different total energy is in the atmospheric response, which in turn affects the evolution of the energy deposition over time. An interesting example is provided by the EB heating run with E$_\textrm{C}$= 5 keV.  We have shown in Sect.~\ref{Sect:3} that if the energy deposited by the 5 keV electrons is high enough, a significant upflow of plasma will start filling the loops, and as a result the electrons streaming from the corona will be stopped at progressively higher heights. On the other hand, if the flows are slower (as in the  E = 10$^{24}$ erg simulations), the electrons can deposit their energy deeper down in the atmosphere for a longer time. As a consequence, the large pressure gradient between the TR and chromosphere, and therefore the strong downflow of \siiv~plasma, will not be observed. In contrast, we find either no shifts or a blueshift (20--30 km s$^{-1}$) in those spectra in the first few 10~s into the simulation, as can be best seen in Fig.~\ref{Fig:sp_si4_1.e24}. 
This result suggests that a moderate blueshift in the \siiv~line can also be found for electron energies as small as 5 keV, if the total energy of the beam is small enough or for certain physical conditions of the loop, as in the case of the  loop with \emph{L}/2 = 100 Mm and T$_{\textrm{LT}}$ = 3MK presented in Sect.~\ref{Sect:6.2}. However, it should be noted that although the 5 keV case for a 10$^{24}$ erg nanoflare also produces a blueshift in the \siiv~line, the main difference between this simulation and the ones with total energy of 6$\cdot$10$^{24}$ erg and larger E$_\textrm{C}$ (10 and 15 keV) will be in the \siiv~intensity, which is at least a factor of10 larger for the latter simulations. 
Figure~\ref{Fig:sp_si4_1.e24} also shows that the higher energy electrons (with cut-off of 10 and 15 keV) are not very efficient at heating the TR. This is because the energy dissipated in the chromosphere can be mostly radiated away without heating the TR and corona. 

Finally, we note that the results of thermal conduction simulations do not change significantly  as a function of total energy deposited during the nanoflare events. For instance, the TR \siiv~spectra in the E = 10$^{24}$ erg simulation will still show a redshift, although on longer timescales than in the  E = 6$\cdot$10$^{24}$ or 10$^{25}$ erg  simulations,  as can be seen in Fig.~\ref{Fig:sp_si4_1.e24}. We note that in all the cases we simulated, in-situ heating of the corona cannot produce a blueshift in the \siiv~line. 
\begin{figure}[!h]
\centering
\includegraphics[width=0.45\textwidth]{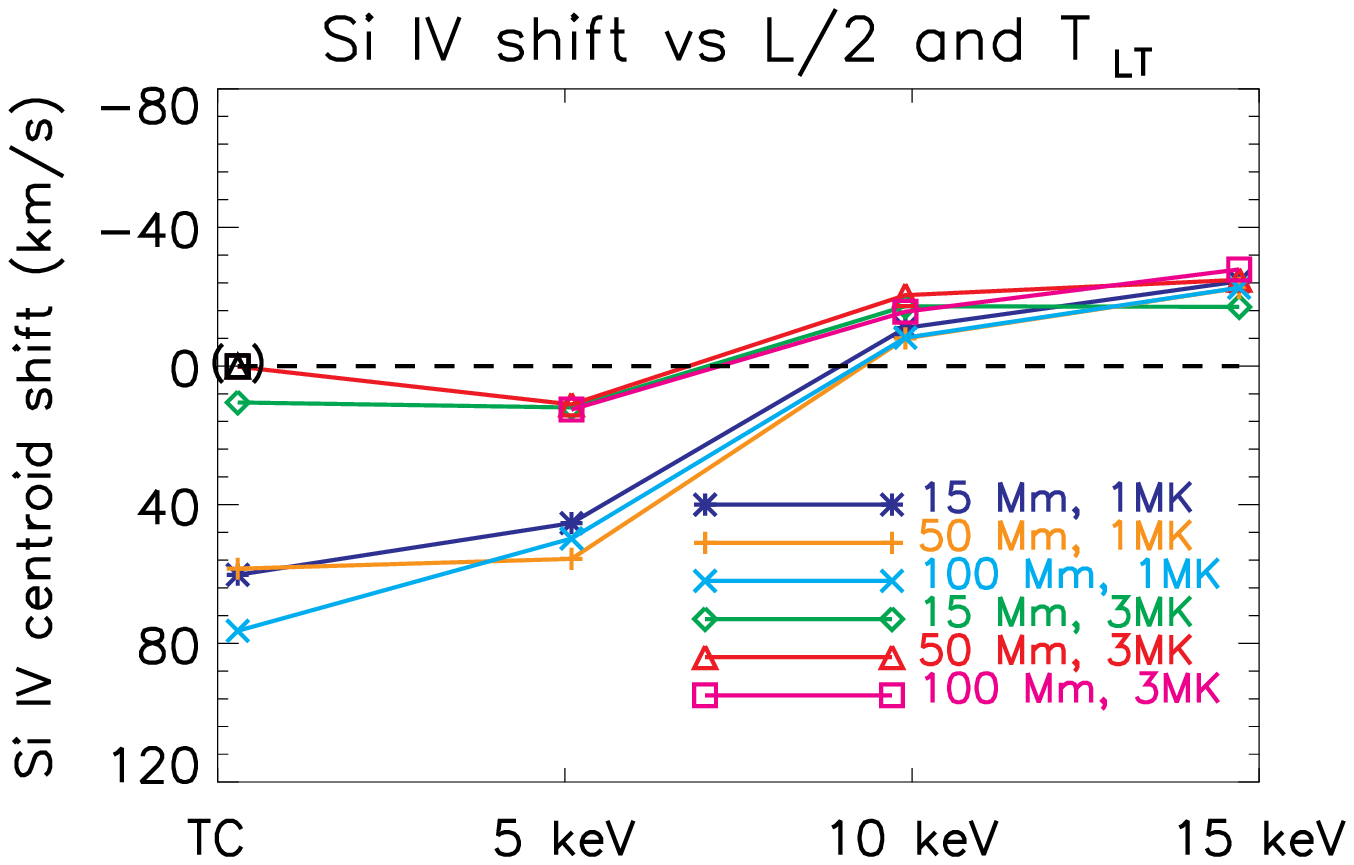}
\includegraphics[width=0.45\textwidth]{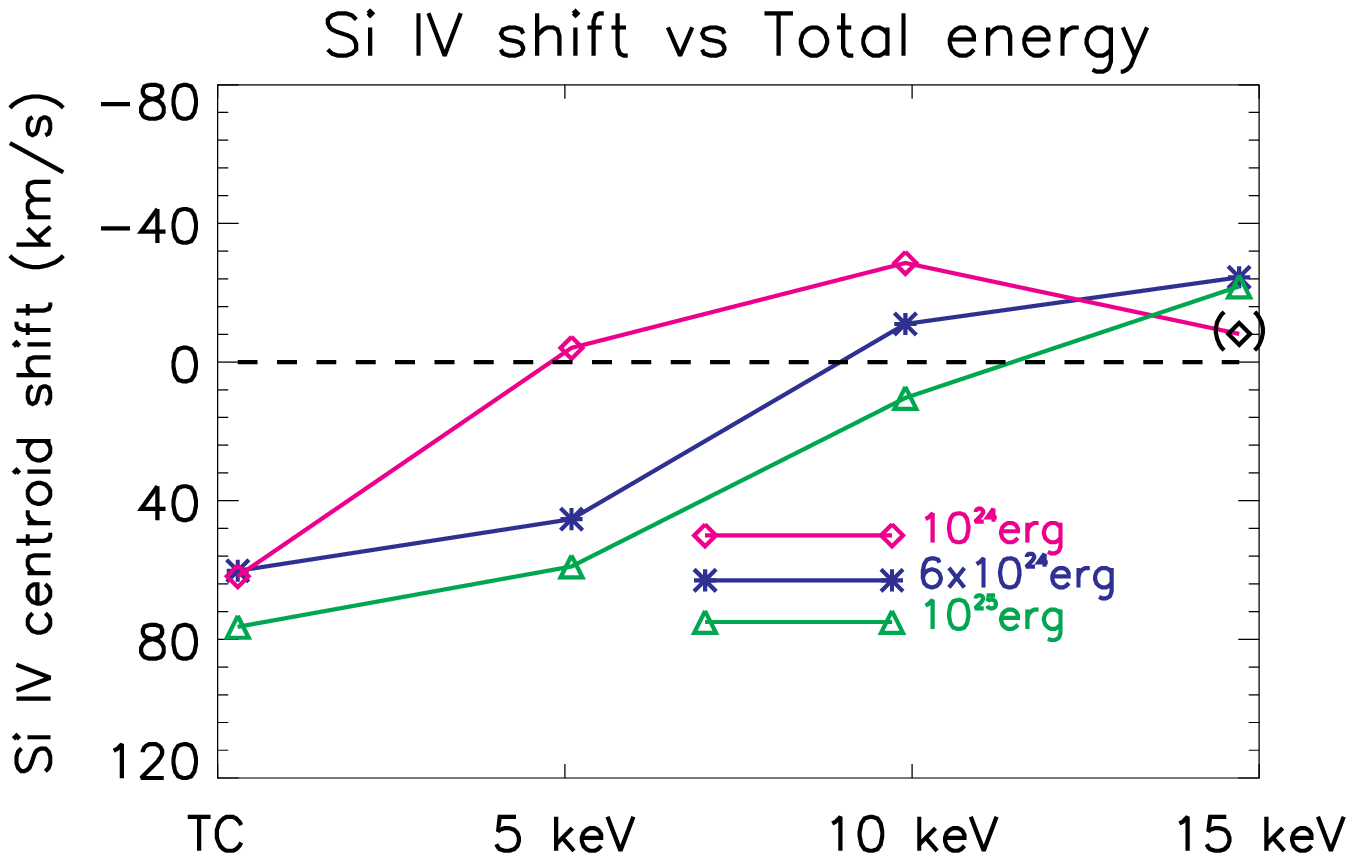}

\caption{Maximum \siiv~Doppler shift velocities obtained for each simulation as a function of different heating models (TC or EB with different values of E$_\textrm{C}$). \emph{Top panel}: Different symbols and colors represent simulations with different initial conditions for the loops (half-loop lengths or temperature) as indicated in the legend. The black symbols in bracket indicate when the line intensity was below 10 DN. \emph{Bottom panel:} Same as the top panel, with different symbols indicating simulations using different total energies.}
\label{Fig:Si4_summary}
\end{figure}

Figure~\ref{Fig:Si4_summary} summarizes the results of \siiv~velocity for the different models we have calculated here and discussed above. The figure shows the maximum \siiv~centroid velocity attained at all times during each simulation as a function of the heating model (thermal conduction or electron beam heating with different electron energy cut-offs) for different initial conditions of the loop (length and temperature, top panel) and different total energies (bottom panel). The black symbols in brackets indicate the cases where the line intensity was too low to be properly detected (below 10 DN).  

The top panel of Fig.~\ref{Fig:Si4_summary} shows that there is an overall trend of \siiv~centroid shifts as a function of heating models, where TC or EB simulations with low E$_\textrm{C}$ present comparable redshifts (or sometimes larger for the TC simulations), while EB simulations tend to show larger blueshifts with increasing E$_\textrm{C}$. The figure also indicates that simulations with different loop lengths and same temperature provide very similar trends, especially for the T$_{\textrm{LT}}$=1MK simulations. We note that the initial density in these loops will be different, because their length is different, as a result of the way the loops are created in equilibrium conditions (as discussed in Sect.~\ref{Sect:2}). However, the density difference is not as large as in loops with the same length but very different temperatures (i.e. 1~and 3~MK loops). On the other hand, simulation runs with different initial loop temperatures (and very different initial densities) produces significantly different results for the TC and EB model with low E$_\textrm{C}$. In addition, if the energy cut-off is large enough, simulations with different initial conditions for the loop produce approximatively the same blueshift in the \siiv~line. 

The bottom panel of Fig.~\ref{Fig:Si4_summary} shows a similar trend of Doppler shifts as a function of heating model for total energies of 6$\cdot$10$^{24}$ or 10$^{25}$ erg. However, this seems not to be the case for simulations with significantly smaller energy (i.e. 10$^{24}$ erg), where even energy distributions with low E$_\textrm{C}$ can cause a small blueshift of the \siiv~line, as also discussed previously.
\section{Conclusions}
\label{Sect:7}
We have carried out an extensive study of nanoflare-heated loops using RADYN 1D HD simulations and investigated the properties of the impulsive heating models and initial plasma conditions, within the parameter space described in Sect. \ref{Sect:2}. This study expands on the initial investigation presented and discussed in T14. Our aim was to reproduce a variety of possible physical scenarios that can be observed in nanoflare heating events, including different heat transport mechanisms (electron beam heating vs in-situ heating of the corona), as well as  different physical initial conditions for the loops (empty vs previously heated loop strands). 
We have chosen to simulate different heating models separately because this approach allows us to isolate the effects of energy deposition in the different cases. In reality, a combination of mechanisms (electron beams with different energies as well as in-situ heating deposition) is most likely to occur (see also discussion in Sect.\ref{Sect:4.2}). 

For our simulations, we used the RADYN code, which provides an accurate modeling of  the chromospheric emission in non-LTE and an advanced treatment of the accelerated electrons through the Fokker-Planck equations.  We have also developed a more realistic initial atmospheric structure including a plage-like chromosphere, based on the work of \cite{Carlsson15}. 

Our numerical investigation has provided several important results: 
\begin{enumerate}
\item The atmospheric response changes significantly as a function of the electron energy cut off, in agreement with the early findings of T14. In particular, lower energy electrons tend to dissipate more energy in the corona, with results similar to the case when the corona is heated in-situ. These electrons also produce very large upflows (a few hundreds km s$^{-1}$), 
in agreement with the prediction made by T14 and \cite{Reep15}, that also nanoflares with low energy electrons could drive "explosive evaporation". In most cases we observe upflows from the TR, as the electrons do not penetrate as deep in the chromosphere as in the flare models, where more energetic electrons are typically assumed. 
Thermal conduction simulations are more effective at heating the loop and produce plasma up to 10--20 MK, which is the range of formation temperatures of highly ionized Fe atoms (such as \fexxi--\fexxiv). The emission of the upflowing hot plasma would however be very faint, because the coronal density is very low ($\approx$~10$^{8}$ cm$^{-3}$) and also the atoms may not be in ionization equilibrium \citep[e.g.][]{Reale08}.
\item The initial conditions of the loops, i.e. temperature and density, are crucial in determining the response of the atmosphere and the \emph{IRIS} and \emph{SDO}/AIA observables. In particular, the same amount of energy deposited by the nanoflare will result in slower flows and a lower intensity of the TR emission for a dense loop (apex density and temperature of $\approx$~10$^{9.5}$ cm$^{-3}$ and 3~MK respectively). If the loop is even hotter and denser, for instance a 5 MK loop with apex density of $\approx$~10$^{10}$ cm$^{-3}$, no significant heating of the TR has been obtained for the nanoflare-sized events we simulated.  Based on our simulations, we can therefore suggest that nanoflare-size heating events in loops with an apex density of more than 10$^{9}$--10$^{9.5}$ cm$^{-3}$ will not  produce any significant TR footpoint brightenings for electron beam heating models with energy cut-off up to 15 keV or thermal conduction models.
\item If the loop plasma has a low density, then the loop length is in many aspects not a crucial parameter:  given the same input parameters for the electron beams and apex temperature, simulations with different loop lengths (and thus slightly different densities, but still low, below 10$^{8}$ cm$^{-3}$ for 50~Mm and 100~Mm loops) will produce qualitatively similar results overall, although the actual values of flows/intensity of the lines might vary.  This does not apply to much denser loops: in this case, the length of the loop can determine different atmospheric responses and TR spectra for the same electron beam model, in particular for the low energy electrons. This is because above a certain density ($\approx$~10$^{9.0}$ cm$^{-3}$) the electrons dissipate their energy more efficiently along the loop and the actual value of density might affect the way this occurs. 
\end{enumerate}
We synthesized the \emph{IRIS} \siiv~1402.77~\AA~and AIA~94~\AA~emission, which can be directly compared with observations.  As pointed out by T14, the analysis of the \siiv~line and its evolution over time provide tight constraints on the possible heating scenarios. We have focused here on the detailed modeling of the optically thin plasma emission, whereas a thorough analysis and discussion of the chromospheric emission from our RADYN simulations is deferred to a future work. 

Our key conclusions are summarized as follows: 

\begin{enumerate}
\item Our single nanoflare loop simulations, where the heating is released in 1~MK and almost empty loops (with apex density of~$\approx$~10$^{8.7}$ cm$^{-3}$), show a sudden increase of \siiv~emission (up to around 4 orders of magnitude) over few seconds, in agreement with the observation of short-lived brightenings typical of the TR \citep[e.g.][T14]{Testa13}. This applies to all the heating models in our numerical investigation, but the maximum \siiv~intensity increase is obtained with the 10 keV electrons, which are more efficient at heating the TR. 
For denser loops (with apex density of ~$\approx$~10$^{9.5}$ cm$^{-3}$), more energetic electrons (10--15 keV) are needed to produce a rapid response of \siiv~emission within the first few seconds of the simulation. For the model runs with 5 keV electrons or in the case of in-situ heating, the \siiv~line becomes bright after only 10~s, as a result of heating by a thermal conduction front from the corona. In any case, the predicted \siiv~change in intensity is also much lower than in the case of a low density loop strand, because the electrons are stopped more efficiently along a denser loop and consequently deposit less energy in the TR.
\item The observation of \siiv~line shift provides a powerful diagnostic of the nanoflare heating properties. For empty loop strands, blueshifts are indicative of higher energy electrons (10--15 keV), whereas simulations with low energy electrons or thermal conduction mostly show \siiv~redshifts. Low energy electrons and in-situ heating of the corona give very similar results and in most of the cases (especially for dense loops) and it is virtually impossible to distinguish the two heating models based on the \emph{IRIS} TR \siiv~line or AIA coronal observables only. 
However, the threshold for TR blueshifts depends mainly on the total nanoflare energy and even 5~keV electrons can produce a \siiv~blueshift in a 10$^{24}$ erg nanoflare.
Combining the observation of \siiv~doppler shift and intensity can help constrain both the energy cut off and energy of the nano flare event. 
Finally, thermal conduction simulations can never produce TR plasma upflows (blueshifts), for all the cases we have analyzed in this work. 
\item Lower energy electrons are more efficient at heating the corona \citep[as also pointed out by ][ and T14]{Reep15}, resulting in more intense AIA~94\AA~emission. In particular, the AIA emission will be higher for loops with higher initial temperature and density in the corona. We also note that our single flare loops predict AIA 94~\AA~intensities much lower than observed. This is not surprising, given that multi-strand loop models, repeated heating events and local energy release in the corona coexisting with accelerated particles might be needed to reproduce the observed coronal emission \citep[e.g.][]{Klimchuk06,Bradshaw12,Reep13a}. Nevertheless, our analysis provides an upper limit for the number of loop strands (of the order of hundreds) needed to match the observational results from AIA. 
\item The chromospheric spectra show significantly different behavior compared with the TR emission. For high E$_\textrm{C}$ ($\geq$~10 keV) the \mgii~intensity is larger for the denser atmosphere than for the initially low density loop. For the low E$_\textrm{C}$ case, no significant increase in \mgii~emission is observed for the dense loop, while more intense, and highly redshifted \mgii~lines, with complex spectral profiles, are predicted for the low density case.

\end{enumerate}
 
The results above suggest a possible scenario of nanoflare heating: (1) first, the intense and short-lived TR brightenings observed at times in the moss by \emph{IRIS} as well as coronal imagers \citep[e.g. \emph{Hi-C}, AIA; e.g. ][]{Testa13} can be explained by impulsive nanoflare heating in initially low density loop strands (our T$_{\textrm{LT}}$=1MK loop model with apex density of $\approx$10$^{8.7}$ cm$^{-3}$); (2) subsequently, while the loops start being filled with high temperature plasma and become denser (our T$_{\textrm{LT}}$=3MK loop model with apex density of $\approx$10$^{9.6}$ cm$^{-3}$), the electrons will no longer be effective at heating the TR plasma anymore, and therefore we will not see a significant increase of emission or shift in the \siiv~spectra. On the other hand, the corona will still be heated to higher temperatures. 

Our work has provided predictions for a large sample of plausible nanoflare heating scenarios, which can be directly compared withe \emph{IRIS} diagnostics of variable footpoint emission associated with coronal heating events in ARs. The findings presented here have also allowed for a better understanding of the heating properties as a function of different input parameter in nanoflare size events.

Finally, we note that the possibility of heating by Alfv\'en wave dissipation in flares has been recently investigated numerically by \cite{Reep16, Kerr16}. Further work will be dedicated to include the Alf\'en waves heating mechanism in our grid of simulations, and compare the predictions from this model with the ones obtained by the electron beam and in-situ heating mechanisms that we have presented in this work.

\begin{acknowledgements}
We thank the referee for the useful comments which helped improve the paper. V.P. and P.T. are supported by NASA grant NNX15AF50G, and by contract 8100002705 from Lockheed-Martin to SAO. P.T. also acknowledges support by NASA grant NNX15AF47G and contract NNM07AB07C to SAO. B.D.P. is supported by NNG09FA40C (IRIS). This research was supported by the Research Council of Norway through its Centres of Excellence scheme, project number 262622, and through grants of computing time from the Programme for Supercomputing. Resources supporting this work were provided by the NASA High-End Computing (HEC) Program through the NASA Advanced Supercomputing (NAS) Division at Ames Research Center. IRIS is a NASA small explorer mission developed and operated by LMSAL with mission operations executed at NASA Ames Research center and major contributions to downlink communications funded by ESA and the Norwegian Space Centre.
CHIANTI is a collaborative project involving researchers at the universities of Cambridge (UK), George Mason and Michigan (USA). The authors thank the International Space Science Institute (ISSI)  for their support and hospitality during the meetings of the ISSI team "New diagnostics of particle acceleration in solar coronal nanoflares from chromospheric observations and modeling".
\end{acknowledgements}

 \bibliographystyle{apj}
\bibliography{v8}
\end{document}